\documentclass[aps,prc,preprint,amsmath,amssymb,showpacs,superscriptaddress]{revtex4}

\usepackage{graphicx}% Include figure files
\usepackage{dcolumn}% Align table columns on decimal point
\usepackage{bm}% bold math
\usepackage{longtable}
\usepackage{color}
\usepackage{CJK}

\begin{document}
%\begin{CJK*}{GBK}{song}
%\preprint{APS/123-QED}

\title{New parametrization for the nuclear covariant energy density functional with point-coupling interaction}

\author{P. W. Zhao }%
\affiliation{State Key Laboratory of Nuclear Physics and Technology, School of Physics, Peking University, Beijing 100871, People's Republic of China}
\author{Z. P. Li }%
\affiliation{State Key Laboratory of Nuclear Physics and Technology, School of Physics, Peking University, Beijing 100871, People's Republic of China}
\author{J. M. Yao }%
\affiliation{School of Physical Science and Technology, Southwest University, Chongqing 400715, People's Republic of China}
\author{J. Meng \footnote{Email: mengj@pku.edu.cn}}%
\affiliation{State Key Laboratory of Nuclear Physics and Technology, School of Physics, Peking University, Beijing 100871, People's Republic of China}
\affiliation{School of Physics and Nuclear Energy Engineering, Beihang University, Beijing 100191, People's Republic of China}
\affiliation{Department of Physics, University of Stellenbosch, Stellenbosch, South Africa}

\date{\today}
\begin{abstract}
A new parametrization PC-PK1 for the nuclear covariant energy
density functional with nonlinear point-coupling interaction is
proposed by fitting to observables of 60 selected spherical nuclei,
including the binding energies, charge radii and empirical pairing
gaps. The success of PC-PK1 is illustrated in the description for
infinite nuclear matter and finite nuclei including the ground-state
and low-lying excited states. Particularly, PC-PK1 provides good
description for isospin dependence of binding energy along either
the isotopic or the isotonic chains, which makes it reliable for
application in exotic nuclei. The predictive power of PC-PK1 is also
illustrated for the nuclear low-lying excitation states in a
five-dimensional collective Hamiltonian in which the parameters are
determined by constrained calculations for triaxial shapes.
\end{abstract}

% 21.60.Jz Nuclear Density Functional Theory and extensions (
%includes Hartree¨CFock and random-phase approximations)
%21.30.Fe Forces in hadronic systems and effective interactions
%21.10.Dr Binding energies and masses
%21.10.Ft Charge distribution

\maketitle

\section{Introduction}
In the past years, the unstable nuclear beams have extended our
knowledge of nuclear physics from the stable nuclei to the unstable
nuclei far from the stability line --- so-called ``exotic nuclei''.
Extensive research in this area shows a lot of entirely unexpected
features and novel aspects of nuclear structure such as the halo
phenomenon~\cite{Tanihata1985PRL,Meng1996PRL,Meng1998PRL}, the
disappearance of traditional magic numbers and the occurrence of new
ones~\cite{Ozawa2000PRL}. The exotic nuclei play important roles in
nuclear astrophysics, since their properties are crucial to
stellar nucleosynthesis. To understand the physics in exotic nuclei,
it becomes very important to find a reliable theory and improve the
reliability for predicting the properties of more exotic nuclei
close to proton and neutron drip lines.

Nuclear energy density functional (EDF) theory~\cite{Bender2003RMP} has played an important role
in a self-consistent description of nuclei. With a few parameters, EDF theory is able
to give a satisfactory description for the ground state properties of spherical and deformed nuclei all over the nuclide chart. Detailed discussion on the EDF theory can be seen in Ref.~\cite{Fayans2000NP} for nonrelativistic representations and in Refs.~\cite{Vretenar2005PRE,Meng2006PPNP} for
relativistic ones.

There exist a number of attractive features in the covariant EDF theory, especially
in its practical applications of self-consistent relativistic mean-field (RMF)
framework~\cite{Vretenar2005PRE,Meng2006PPNP}. The most obvious one is the natural
inclusion of the nucleon spin degree of freedom and the resulting nuclear spin-orbit
potential that emerges automatically with the empirical strength in a covariant way.
The relativistic effects are responsible for the empirical existence of approximate
pseudospin symmetry in the nuclear single-particle spectra~\cite{Ginocchio2005PRE}.
Moreover, a covariant treatment of nuclear matter provides a distinction between scalar
and four-vector nucleon self energies, leading to a natural saturation mechanism.

The most widely used RMF framework is based on the finite-range meson-exchange representation
(RMF-FR), in which the nucleus is described as a system of Dirac nucleons which interact with
each other via the exchange of mesons. The isoscalar-scalar $\sigma$ meson, the
isoscalar-vector $\omega$ meson, and the isovector-vector $\rho$ meson build the minimal set of meson fields that, together with the electromagnetic field, is necessary for a description of bulk and single-particle nuclear properties. Moreover, a quantitative treatment of nuclear matter and finite nuclei needs a medium dependence of effective mean-field interactions, which can be introduced by including nonlinear meson self-interaction terms in the Lagrangian or by assuming explicit density dependence for the meson-nucleon couplings. Of course, at the energy characteristic for nuclear binding and low-lying excited states, the heavy-meson exchange ($\sigma$, $\omega$, $\rho$) is just a convenient representation of the effective nuclear interaction.

Since the exchange of heavy mesons is associated with short-distance dynamics that cannot be resolved
at low energies, as an alternative, the relativistic point-coupling (RMF-PC) model~\cite{Nikolaus1992PR,Burvenich2002PR} is proposed by
using the zero-range point-coupling interaction instead of the meson exchange, i.e., in each channel
(scalar-isoscalar, vector-isoscalar, scalar-isovector, and vector-isovector) meson exchange is replaced by the corresponding local four-point (contact) interaction between nucleons. Analogously, in the case of contact interactions, the medium effects can be taken into account by including higher-order (nonlinear coupling) interaction terms or by assuming a density dependence of strength parameters for the coupling interactions.

In recent years, the RMF-PC model has attracted more and more
attentions due to the following advantages. Firstly, it avoids the
possible physical constrains introduced by explicit usage of the
Klein-Gordon equation to describe mean meson fields, especially the
fictitious $\sigma$ meson. Secondly, it is possible to study the
role of naturalness~\cite{Friar1996PR,Manohar1984NP} in effective
theories for nuclear structure related problems. Thirdly, it provides
more opportunities to investigate its relationship to the
nonrelativistic approaches~\cite{Sulaksono2003ANNP}. Finally, it is
relatively easy to study the effects beyond mean-field for the nuclear low-lying collective excited states.

In practical application of the RMF-PC model, the most widely
used nonlinear coupling parameterizations include
PC-LA~\cite{Nikolaus1992PR} and PC-F1~\cite{Burvenich2002PR}. PC-LA
is determined by the ground-state observables of $^{16}\rm O$,
$^{88}\rm Sr$, and $^{208}\rm Pb$. Due to the explicit omission of
the pairing interaction, the pairing effects are not included in the
fitting procedure. Moreover, the test for naturalness in
Ref.~\cite{Friar1996PR} shows that only six of the nine coupling
constants are natural. As an improvement, PC-F1 is optimized to observables of 17 spherical nuclei including open-shell nuclei,
and the pairing correlation is considered through a standard BCS
approach in the fitting procedure. Furthermore, all the coupling
constants of PC-F1 are turned out to be
natural~\cite{Burvenich2002PR}. However, the isospin
dependence of binding energy given by PC-F1 along either the isotopic or the isotonic chains deviates from the data remarkably.

Recently, a density-dependent parametrization DD-PC1 is proposed from the equation of state (EOS) of nuclear matter and
the masses of 64 axially deformed nuclei in the mass regions $A \simeq 150-180$ and $A \simeq
230-250$~\cite{Niksic2008PR}. Although it reproduces the binding
energies, deformations, and charge radii of deformed nuclei quite
well, the differences between the predicted binding energies and the
corresponding data are somewhat large for spherical nuclei.

Therefore, it is necessary to have a new parametrization for the nuclear covariant energy density functional with point-coupling
interaction to describe both the nuclear matter and finite nuclei
properties. In this work, a new parametrization PC-PK1 with
nonlinear coupling interactions is proposed. In Sec. II, the
theoretical framework for the relativistic point-coupling model is
briefly outlined. The numerical details are given in Sec. III. In
Sec. IV-VII, a series of illustrative descriptions for the nuclear
matter, spherical nuclei, deformed nuclei as well as the nuclear
excited properties are presented. Finally, a summary is given in
Sec. VIII.

\section{Theoretical framework}
The basic building blocks of RMF theory with point-coupling vertices are
\begin{equation}
    (\bar\psi{\cal O}\Gamma\psi),~~~~~{\cal O}\in\{1,\vec{\tau}\},~~~~~\Gamma\in\{1,\gamma_\mu,\gamma_5,
    \gamma_5\gamma_\mu,\sigma_{\mu\nu}\},
\end{equation}
where $\psi$ is Dirac spinor field of nucleon, $\vec{\tau}$ is the isospin Pauli matrix, and $\Gamma$
generally denotes the $4\times4$ Dirac matrices. There are ten such building blocks characterized by
their transformation characteristics in isospin and Minkowski space. In this paper, vectors in
the isospin space are denoted by arrows and the space vectors by bold type. Greek indices $\mu$ and
$\nu$ run over the Minkowski indices $0$, $1$, $2$, and $3$.

A general effective Lagrangian can be written as a power series in
$\bar\psi{\cal O}\Gamma\psi$ and their derivatives. We start with
the following Lagrangian density of the point-coupling model
\begin{equation}\label{EQ:LAG}
  {\cal L} = {\cal L}^{\rm free}+{\cal L}^{\rm 4f}+{\cal L}^{\rm hot}+{\cal L}^{\rm der}+{\cal L}^{\rm em},
\end{equation}
which is divided as the Lagrangian density for free nucleons ${\cal
L}^{\rm free}$,
\begin{equation}\label{EQ:LAGfree}
{\cal L}^{\rm free}=\bar\psi(i\gamma_\mu\partial^\mu-m)\psi,
\end{equation}
the four-fermion point-coupling terms ${\cal L}^{\rm 4f}$,
\begin{eqnarray}
{\cal L}^{\rm 4f}&=&-\frac{1}{2}\alpha_S(\bar\psi\psi)(\bar\psi\psi)
-\frac{1}{2}\alpha_V(\bar\psi\gamma_\mu\psi)(\bar\psi\gamma^\mu\psi)\nonumber\\
&&-\frac{1}{2}\alpha_{TS}(\bar\psi\vec{\tau}\psi)(\bar\psi\vec{\tau}\psi)
-\frac{1}{2}\alpha_{TV}(\bar\psi\vec{\tau}\gamma_\mu\psi)(\bar\psi\vec{\tau}\gamma^\mu\psi),
\end{eqnarray}
the higher order terms ${\cal L}^{\rm hot}$ which are responsible for
the effects of medium dependence,
\begin{equation}
{\cal L}^{\rm hot}=-\frac{1}{3}\beta_S(\bar\psi\psi)^3-\frac{1}{4}\gamma_S(\bar\psi\psi)^4-\frac{1}{4}
    \gamma_V[(\bar\psi\gamma_\mu\psi)(\bar\psi\gamma^\mu\psi)]^2,
\end{equation}
the gradient terms ${\cal L}^{\rm der}$ which are included to simulate the effects of
finite-range,
\begin{eqnarray}
{\cal L}^{\rm der} &=& -\frac{1}{2}\delta_S\partial_\nu(\bar\psi\psi)\partial^\nu(\bar\psi\psi)
-\frac{1}{2}\delta_V\partial_\nu(\bar\psi\gamma_\mu\psi)\partial^\nu(\bar\psi\gamma^\mu\psi) \nonumber \\
&&-\frac{1}{2}\delta_{TS}\partial_\nu(\bar\psi\vec\tau\psi)\partial^\nu(\bar\psi\vec\tau\psi)
-\frac{1}{2}\delta_{TV}\partial_\nu(\bar\psi\vec\tau\gamma_\mu\psi)\partial^\nu(\bar\psi\vec\tau\gamma_\mu\psi),
\end{eqnarray}
and the electromagnetic interaction terms ${\cal L}^{\rm em}$,
\begin{equation}\label{EQ:LAGelecmag}
{\cal L}^{\rm em}=-\frac{1}{4}F^{\mu\nu}F_{\mu\nu}-e\frac{1-\tau_3}{2}\bar\psi\gamma^\mu\psi A_\mu.
\end{equation}

For the Lagrangian density in Eq.~(\ref{EQ:LAG}), $m$ is the nucleon
mass and $e$ is the charge unit for protons. $A_\mu$ and
$F_{\mu\nu}$ are respectively the four-vector potential and field
strength tensor of the electromagnetic field. There are totally 11
coupling constants, $\alpha_S$, $\alpha_V$, $\alpha_{TS}$,
$\alpha_{TV}$, $\beta_S$, $\gamma_S$, $\gamma_V$, $\delta_S$,
$\delta_V$, $\delta_{TS}$, and $\delta_{TV}$, in which $\alpha$
refers to the four-fermion term, $\beta$ and $\gamma$ respectively
the third- and fourth-order terms, and $\delta$ the derivative
couplings. The subscripts $S$, $V$, and $T$ respectively indicate
the symmetries of the couplings, i.e., $S$ stands for scalar, $V$
for vector, and $T$ for isovector.

From former experience~\cite{Burvenich2002PR}, we neglect the isovector-scalar channel in
Eq.~(\ref{EQ:LAG}) since a fit including the isovector-scalar
interaction does not improve the description of nuclear ground-state
properties. Consequently, there are nine free parameters in the present
RMF-PC model, which are comparable with those in the RMF-FR model.
Furthermore, the pseudoscalar $\gamma_5$ and pseudovector
$\gamma_5\gamma_\mu$ channels are also neglected in Eq.~(\ref{EQ:LAG}) since
they do not contribute at the Hartree level due to parity
conservation in nuclei.

Similar to the RMF-FR case, the mean-field approximation leads to
the replacement of the operators $\bar\psi(\hat{{\cal
O}}\Gamma)_i\psi$ in Eq. (\ref{EQ:LAG}) by their expectation values
which become bilinear forms of the nucleon Dirac spinor $\psi_k$,
\begin{equation}
  \bar\psi(\hat{{\cal O}}\Gamma)_i\psi\rightarrow\langle\Phi|\bar\psi(\hat{{\cal O}}\Gamma)_i\psi|\Phi\rangle
  =\sum\limits_kv_k^2\bar\psi_k(\hat{{\cal O}}\Gamma)_i\psi_k,
\end{equation}
where $i$ indicates $S$, $V$, and $TV$. The sum $\sum$ runs over only
positive energy states with the occupation probabilities $v_k^2$, i.e., the ``no-sea'' approximation.
Based on these approximations, one finds the energy density functional for a nuclear system
\begin{equation} \label{Energy}
 E_{\rm DF}[{\bm{\tau}}, \rho_S, j^\mu_i, A_\mu] = \int d^3r~{\mathcal{E}(\bm{r}}),
\end{equation}
with the energy density
\begin{equation}
 \mathcal{E}(\bm{r})= \mathcal{E}^{\rm kin}(\bm{r})
                  +  \mathcal{E}^{\rm int}(\bm{r})
                  +  \mathcal{E}^{\rm em}(\bm{r}),
\end{equation}
which is composed of a kinetic part
\begin{equation}
 \mathcal{E}^{\rm kin}(\bm{r}) = \sum_k\,v_k^2~{\psi^\dagger_k (\bm{r})
    \left(\bm{\alpha}\cdot\bm{p} + \beta m\right )\psi_k(\bm{r})},
\end{equation}
an interaction part
\begin{eqnarray}
 \mathcal{E}^{\rm int}(\bm{r})
 &=& \frac{\alpha_S}{2}\rho_S^2+\frac{\beta_S}{3}\rho_S^3
    + \frac{\gamma_S}{4}\rho_S^4+\frac{\delta_S}{2}\rho_S\triangle\rho_S\nonumber \\
 &&+ \frac{\alpha_V}{2}j_\mu j^\mu + \frac{\gamma_V}{4}(j_\mu j^\mu)^2 +
       \frac{\delta_V}{2}j_\mu\triangle j^\mu  \\
 && +  \frac{\alpha_{TV}}{2}\vec j^{\mu}_{TV}\cdot(\vec j_{TV})_\mu+\frac{\delta_{TV}}{2}
    \vec j^\mu_{TV}\cdot\triangle(\vec j_{TV})_{\mu},\nonumber
\end{eqnarray}
with the local densities and currents
\begin{subequations}\label{currents}
\begin{eqnarray}
  \rho_S(\bm{r})          &=&\sum_{k }v^2_k\bar\psi_k(\bm{r})\psi_k(\bm{r}),\\
  j^\mu_{V}(\bm{r})       &=&\sum_{k }v^2_k\bar\psi_k(\bm{r})\gamma^\mu\psi_k(\bm{r}),\\
  \vec j^\mu_{TV}(\bm{r}) &=&\sum_{k }v^2_k\bar\psi_k(\bm{r})\vec\tau\gamma^\mu\psi_k(\bm{r}),
\end{eqnarray}
\end{subequations}
and an electromagnetic part
\begin{equation}
 \mathcal{E}^{\rm em}(\bm{r})=\frac{1}{4}F_{\mu\nu}F^{\mu\nu}-F^{0\mu}\partial_0A_\mu+eA_\mu j^\mu_p.
\end{equation}

Minimizing the energy density functional Eq.~(\ref{Energy}) with
respect to $\bar\psi_k$, one obtains the Dirac equation for the
single nucleons
\begin{equation}
 \label{DiracEq}
   [\gamma_\mu(i\partial^\mu-V^\mu)-(m+S)]\psi_k=0.
\end{equation}
The single-particle effective Hamiltonian contains local scalar $S(\bm{r})$ and vector
$V^\mu(\bm{r})$ potentials,
\begin{equation}
\label{potential}
  S(\bm{r})    =\Sigma_S, \quad
  V^\mu(\bm{r})=\Sigma^\mu+\vec\tau\cdot\vec\Sigma^\mu_{TV},
\end{equation}
where the nucleon scalar-isoscalar $\Sigma_S$, vector-isoscalar $\Sigma^\mu$, and
vector-isovector $\vec\Sigma^\mu_{TV}$ self-energies are given in terms of the various
densities,
\begin{subequations}
\begin{eqnarray}
 \Sigma_S           &=&\alpha_S\rho_S+\beta_S\rho^2_S+\gamma_S\rho^3_S+\delta_S\triangle\rho_S,\\
 \Sigma^\mu         &=&\alpha_Vj^\mu_V +\gamma_V (j^\mu_V)^3
                      +\delta_V\triangle j^\mu_V + e A^\mu,\\
 \vec\Sigma^\mu_{TV}&=& \alpha_{TV}\vec j^\mu_{TV}+\delta_{TV}\triangle\vec j^\mu_{TV}.
\end{eqnarray}
\end{subequations}
For a system with time reversal invariance, the space-like
components of the currents $\bm{j}_i$ in Eq.~(\ref{currents}) and the vector potential $\bm{V}(\bm{r})$ in Eq.~(\ref{potential})
vanish. Furthermore, one can assume that the nucleon single-particle
states do not mix isospin, i.e., the single-particle states are
eigenstates of $\tau_3$. Therefore only the third component of
isovector potentials $\vec\Sigma^\mu_{TV}$ survives. The Coulomb
field $A_0$ is determined by Poisson's equation.

In addition to the self-consistent mean-field potentials, for
open-shell nuclei, pairing correlations are taken into account by
the BCS method with a smooth cutoff factor $f_k$ to simulate the
effects of finite-range~\cite{Krieger1990NP,Bender2000EPJ}, i.e., we
have to add to the functional Eq.~(\ref{Energy}) a pairing energy
term of the form depending on the pairing tensor $\kappa$,
\begin{equation}
\label{PairingE} E_{\rm pair}[\kappa,\kappa^*] = \sum_{kk'>0}
f_{k^{}} f_{k'}\langle k {\bar k} \vert V^{pp} \vert k'{\bar
k}'\rangle \kappa^\ast_{k} \kappa_{k'},
\end{equation}
with the smooth cut-off weight factor
\begin{equation}
\label{Weight} f_k  =\frac{1}{1+\exp[( \epsilon_k- \epsilon_F-\Delta
E_\tau)/\mu_\tau]},
\end{equation}
where $\epsilon_k$ is the eigenvalue of the self-consistent
single-particle field, and $\epsilon_F$ is the chemical potential
determined by the particle number, $\langle \Phi\vert \hat N_\tau
\vert \Phi \rangle = N_\tau$, with $N_\tau$ the particle number of
neutron or proton. The cut-off parameters $\Delta E_\tau$ and
$\mu_\tau=\Delta E_\tau/10$ are chosen in such a way that
$2\displaystyle\sum_{k>0}f_k = N_\tau
+1.65N^{2/3}_\tau$~\cite{Bender2000EPJ}.

In the following calculations, a density-independent $\delta$-force in the pairing channel
is adopted. Thus, the pairing energy is given by
\begin{equation}
\label{PairingDFT}
E_{\rm pair}[\kappa,\kappa^*]=
-\sum_{\tau=n,p} \dfrac{V_\tau}{4}\int d^3r\kappa^\ast_\tau(\bm{r}) \kappa_\tau(\bm{r}),
\end{equation}
where $V_\tau$ is the constant pairing strength and the
pairing tensor $\kappa(\bm{r})$ reads
\begin{equation}
 \kappa(\bm{r})
 =-2\sum_{k>0}f_ku_kv_k\vert\psi_k(\bm{r})\vert^2.
\end{equation}
The pairing strength parameters $V_{\tau}$ can be adjusted by
fitting the average single-particle pairing gap
\begin{equation}
\langle\Delta\rangle\equiv%
\frac{\sum_{k}f_{k}u_kv_{k}\Delta_{k}}{\sum_{k}f_{k}u_kv_{k}}
\label{avgap}%
\end{equation}
to the data obtained with a five-point formula.

As the translational symmetry is broken in the mean-field approximation,
proper treatment of center-of-mass (c.m.) motion is very
important and here the c.m. correction energy is calculated by microscopic c.m. correction
\begin{equation}
 \label{Eq:Ecm}
 E^{\rm mic}_{\rm c.m.}=-\frac{1}{2mA}\langle\hat{\bm P}^{2}_{\rm c.m.}\rangle,
\end{equation}
with $A$ mass number and $\hat{\bm{P}}_{\rm c.m.}=\sum_i^A
\hat{\bm{p}}_i$ the total momentum in the c.m. frame. It has been
shown that the microscopic c.m. correction provides more reasonable
and reliable results than phenomenological
ones~\cite{Bender2000EPJa,Long2004PR,Zhao2009CPL}.

Therefore, the total energy for the nuclear system becomes
\begin{equation}
E_{\rm tot}
 = E_{\rm DF}[\bm{\tau}, \rho_S, j^\mu_i, A_\mu]
  +E_{\rm pair}[\kappa,\kappa^*] +E^{\rm mic}_{\rm c.m.}.
\end{equation}

\section{Numerical details}

In this work, a series of calculations have been performed for both
the spherical and deformed nuclei. The Dirac equation for nucleons is
solved in a three-dimensional harmonic oscillator basis~\cite{Gambhir1990Ann.Phys.(N.Y.)}.
For spherical calculations, by increasing the fermionic shells from $N_f=20$ to $N_f=22$, the binding energy,
charge radius, and neutron skin thickness in $^{208}\rm Pb$ change by 0.003\%, 0.007\%, and 0.1\% respectively.
For $^{240}\rm U$, the binding energy
changes by 0.001\% from the axially deformed calculation with
$N_f=16$ to $N_f=18$. Therefore, a basis of 20 major oscillator
shells is used in the spherical calculations and 16 shells in the
axially deformed cases. The triaxial calculations are performed with
$N_f=12$, which, for Nd isotopes, provides an accuracy of 0.04\% for
binding energies in comparison with the calculations with $N_f=14$.
To achieve an accuracy of $\sim 100$ keV in the description of both the fission
barrier and the energy of the isomer state in $^{240}\rm Pu$, a basis of 20
oscillator shells has been adopted in both the axial and triaxial calculations.

In order to determine the parameters of Lagrangian density in
Eq.~(\ref{EQ:LAG}) and the pairing strength in Eq.~(\ref{PairingDFT}), a multiparameter
fitting to both the binding energies and charge radii for selected
spherical nuclei is performed with the Levenberg-Marquardt
method~\cite{Press1992B}. As usual, the masses of neutron and proton
are fixed as 939 MeV. The corresponding data~\cite{Audi2003NPa, DeVries1987ADNDT, Nadjakov1994ADNDT} for selected spherical
nuclei used in the fitting procedure are listed in
Table~\ref{Tabel:mass} and~\ref{Tabel:rch}. The empirical neutron
pairing gaps for $^{122}\rm Sn$, $^{124}\rm Sn$, and $^{200}\rm Pb$
as well as the proton ones for $ ^{92}\rm Mo$, $^{136}\rm Xe$, and
$^{144}\rm Sm$ obtained with five-point formula are also employed to
constrain the pairing strengths.

With the experimental observable $O_i^{\rm exp}$ and the calculated
value $O_i^{\rm cal}$, by minimizing the square deviation
\begin{equation}
  \chi^2({\bf a})=\sum\limits_i^N\left[\frac{O_i^{\rm exp}-O_i^{\rm cal}({\bf a})}{\omega_i}\right]^2,
\end{equation}
the ensemble of parameters $\bf a$ can be obtained. Furthermore, in order to balance the influence of different observables, the weight $\omega_i$ is introduced for binding energies, charge radii, and empirical pairing gaps respectively. The corresponding weight $\omega_i$ is roughly determined by the desired accuracy. Here the weights $\omega_i$ are respectively $1.00$ MeV for binding energies, $0.02$ fm for charge radii, and $0.05$ MeV for empirical pairing gaps.
A new parameter set PC-PK1, which contains the nine
coupling constants in Eq.~(\ref{EQ:LAG}) and the pairing strength in
Eq.~(\ref{PairingDFT}), is obtained and listed in
Table~\ref{Tabel:PC-PK1}.

By scaling the coupling constants in accordance with the QCD-based
Lagrangian, the naturalness in effective theories can be investigated~\cite{Friar1996PR,Manohar1984NP}. According to the
QCD-based Lagrangian~\cite{Manohar1984NP},
\begin{equation}
  {\cal L}\sim-c_{ln}\left[\frac{\bar\psi\psi}{f^2_\pi\Lambda}\right]^l\left[\frac{\partial^\mu}{\Lambda}\right]^n
  f^2_\pi\Lambda^2,
\end{equation}
with $\psi$ the nucleon field, $f_\pi=92.5~\rm MeV$ the pion decay
constant, and $\Lambda=770~\rm MeV$ a generic QCD large-mass scale
respectively, by taking into account the role of chiral symmetry in
weakening $N$-body forces by
$\Delta=l+n-2\geqslant0$~\cite{Weinberg1979Physica,Weinberg1990PL},
it has been found that six of the nine coupling constants in PC-LA
and all of them in PC-F1 are natural, i.e., the QCD-scaled coupling
constants $c_{ln}$ are of order
unity~\cite{Friar1996PR,Burvenich2002PR}.

Similarly, the nine coupling constants of PC-PK1 are also tested for
the naturalness and all the dimensionless coefficients $c_{ln}$ are
of order $1$, as shown in the last column of Table~\ref{Tabel:PC-PK1}, which indicates that all the coupling constants in PC-PK1 are natural.

Tables~\ref{Tabel:mass} and \ref{Tabel:rch} list respectively the
binding energies and charge radii for nuclei selected in the
determination of PC-PK1, PC-F1, PC-LA, and NL3*~\cite{Lalazissis2009PL} effective
interactions. The corresponding root mean square (rms) deviation
$\Delta$ together with the root of relative square (rrs) deviation $\delta$
for the binding energy and charge radius are given in the last two
rows of Table~\ref{Tabel:mass} and \ref{Tabel:rch}, respectively. Compared with
the other effective interactions, the newly obtained PC-PK1 provides a
much better description for the experimental binding energies and the same good description for the charge radii.

\section{Nuclear matter properties}

In this section, we will present the saturation properties and the equation of state (EOS) for
nuclear matter in the covariant EDF with PC-PK1. The results will be compared with
the corresponding empirical values as well as the predictions with
PC-LA, PC-F1, DD-PC1, NL3*, and PK1~\cite{Long2004PR}.

\subsection{Saturation properties}

The saturation properties, including the binding energy per nucleon
$E/A$, saturation density $\rho_0$, incompressibility $K_0$, nucleon
effective mass $M^\ast_D$ and $M^\ast_L$, symmetry energy $E_{\rm
sym}$, as well as the characteristics $L$ and $K_{\rm asy}$ for the
density dependence of $E_{\rm sym}$ will be investigated.

There are several kinds of nucleon effective mass~\cite{Jaminon1989PR,Dalen2005PRL}. Here we mainly
focus on the Dirac mass and Landau mass. The Dirac mass $M_D^\ast$
is defined through the nucleon scalar self-energy in the Dirac
equation, i.e., $M_D^\ast=M+\Sigma_S$. It is directly related to the
spin-orbit potential in finite nuclei and is thus a genuine
relativistic quantity without nonrelativistic correspondence.
While the Landau mass $M^\ast_L=pdp/dE$ is related to the density of state both in relativistic and non-relativistic models. In relativistic models, the relation between the Dirac mass and the Landau mass is $M^\ast_L = \sqrt{p_F^2+M^{\ast2}_D} $, where $p_F$ is the Fermi momentum.

The density dependence of the nuclear symmetry energy is very
important to understand the properties of exotic nuclei with extreme
isospin values, in particular the slope $L\equiv3\rho_0(dE_{\rm
sym}/d\rho)_{\rho=\rho_0}$ and curvature $K_{\rm
sym}\equiv9\rho_0^2(d^2E_{\rm sym}/d^2 \rho)_{\rho=\rho_0}$ of the
symmetry energy at the saturation density $\rho_0$. In
Refs.~\cite{Prakash1985PR,Baran2002NP}, the isospin-dependent part,
$K_{\rm asy}\approx K_{\rm sym}-6L$, in the isobaric
incompressibility $K(\delta)=K_0+K_{\rm asy}\delta^2$ (with
$\delta\equiv(\rho_n-\rho_p)/\rho$), is often used to characterize
the density dependence of the symmetry energy as both $L$
and $K_{\rm asy}$ can be extracted from the experiment empirically
(see Ref.~\cite{Chen2007PR} and references therein).

In Table~\ref{Tabel:nuclearmatter}, the saturation properties for
nuclear matter, including the binding energy per nucleon
$E/A$, saturation density $\rho_0$, incompressibility $K_0$, nucleon
effective mass $M^\ast_D$ and $M^\ast_L$, symmetry energy $E_{\rm
sym}$, as well as the characteristics $L$ and $K_{\rm asy}$ for the
density dependence of $E_{\rm sym}$, predicted by PC-PK1 are listed in
comparison with those by both point-coupling DD-PC1, PC-F1, PC-LA and meson
exchange NL3*, PK1 sets. In general, PC-PK1 gives good description
for the saturation properties of nuclear matter.
In particular, the predicted values for binding energy per nucleon and density at the saturation point are $-16.12~\rm MeV$ and $0.154~\rm fm^{-3}$, which agree well with the empirical values $-16\pm1~\rm MeV$ and $0.166\pm0.018~\rm fm^{-3}$~\cite{Brockmann1990PR}, respectively. Moreover, the incompressibility given by PC-PK1 is $238$~MeV.

For the effective masses, all the effective interactions give
reasonable values between $0.55$ and $0.60$ for the Dirac mass
$M_D^\ast/M$~\cite{Marketin2007PR} as required by the spin-orbit
splitting data in finite nuclei, but smaller Landau masses
$M_L^\ast/M$ compared with the empirical constraint
$0.8\pm0.1$~\cite{Reinhard1999NP}, implying that they would give a
small single-particle level density at the Fermi energy in finite
nuclei as compared with data.

The symmetry energies in the calculations with the nonlinear effective interactions PC-F1, PC-LA, NL3* and PK1 are always larger than the empirical value (around 32 MeV) by around
16-18\%, which is reduced to 11\% for PC-PK1. Therefore all the interactions would predict large neutron skin thicknesses in
finite nuclei except DD-PC1, which is adjusted by fixing $E_{\rm sym}=33~\rm MeV$. Moreover, the empirical $L$ ($88\pm25 ~\rm MeV$)~\cite{Chen2007PR}
and $K_{\rm asy}$ ($-550\pm100 ~\rm MeV$)~\cite{Li2007PRL} have been
reproduced quite well by PC-PK1.

\subsection{Equation of state}

\begin{figure}[!htbp]
\includegraphics[width=8cm]{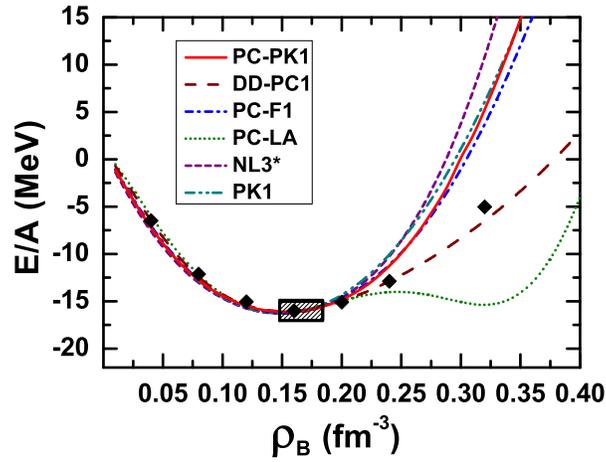}
\caption{(Color online) The binding energy per nucleon $E/A$ for nuclear matter as a function of the
baryon density
$\rho_B$ given by PC-PK1, DD-PC1, PC-F1, PC-LA, NL3*, and PK1. The shaded area indicates the empirical
value~\cite{Brockmann1990PR} and the filled diamonds present the microscopic results of
the ab-initio variational calculation~\cite{Akmal1998PR}.}
\label{fig:nuclearmatter}
\end{figure}

In Fig.~\ref{fig:nuclearmatter}, the binding energy per nucleon
$E/A$ for nuclear matter as a function of the baryon density
$\rho_B$ given by PC-PK1 is shown in comparison with those by
DD-PC1, PC-F1, PC-LA, NL3*, and PK1. All the effective interactions
predict the similar $E/A$ behavior with density below $\rho_B=0.20$
fm$^{-3}$ due to the constraints from the properties of the finite
nuclei. Divergence appears at supra-saturation densities, especially
for the results given by PC-LA. It implies that the properties of
finite nuclei are not sufficient in the determination of EDF to
describe the EOS at supra-saturation densities that are directly
related to the maximal mass of neutron star. The prediction by
PC-PK1 is consistent with those by PK1 and PC-F1, while softer than
those by NL3*. The ab-initio variational calculation for the
symmetric nuclear matter~\cite{Akmal1998PR} is also given for
comparison, which coincides with the relativistic EOS with density
below $\rho_B=0.20$ fm$^{-3}$ but predicts softer EOS behavior at
supra-saturation densities than the relativistic ones, except those
given by PC-LA and DD-PC1. One should note that DD-PC1 has been
adjusted to the EOS given by ab-initio variational
calculations~\cite{Niksic2008PR}.

\section{Spherical Nuclei}

In this section, we will present the binding
energies, two-neutron separation energies, single-particle levels,
charge radii and neutron skin thicknesses for selected spherical
isotopes and isotones in different mass regions in the covariant EDF with PC-PK1. The results will be compared with the
corresponding data available as well as the predictions with DD-PC1, PC-F1,
PC-LA, and NL3* sets.

\subsection{Binding energy}

\begin{figure*}[!htbp]
\includegraphics[width=12cm]{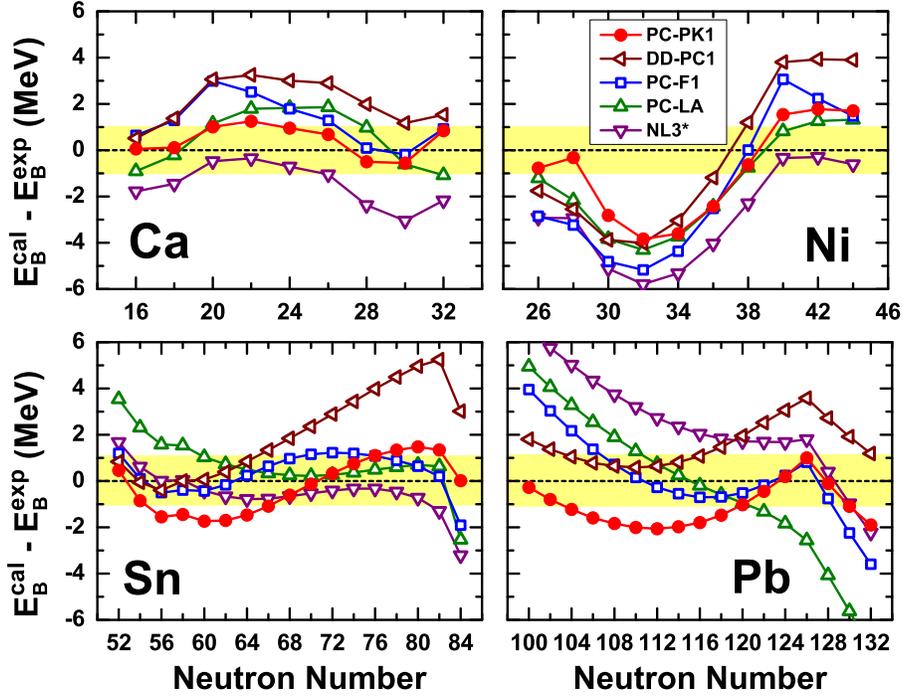}
\caption{(Color online) Deviations of the calculated binding energies for Ca, Ni, Sn, and Pb
isotopes by PC-PK1 from the data~\cite{Audi2003NPa} in comparison with those by DD-PC1,
PC-F1, PC-LA, and NL3*.}
\label{fig:sph_isotope}
\end{figure*}

The binding energies for the Ca, Ni, Sn, and Pb isotopes are
calculated with PC-PK1 and their deviations from the
data~\cite{Audi2003NPa} are shown in Fig.~\ref{fig:sph_isotope} in
comparison with those with DD-PC1, PC-F1, PC-LA, and NL3*. The calculation with
PC-PK1 reproduces the experimental binding energies within $1$ MeV
for the Ca isotopes and $2$ MeV for both Sn and Pb isotopes. For Ni
isotopes, although remarkable improvement is achieved by PC-PK1 in
comparison with results by the other interactions, there is still an
underestimation of $2.4-5.2$ MeV for the even-even $^{58-64}\rm Ni$.
Former investigations have shown that these nuclei are soft against
deformation~\cite{Hilaire2007EJP}. Therefore, the dynamic
correlation energies gained by restoration of rotational symmetry
and configuration mixing are expected to reduce these deviations.
Similarly, the underestimations of the binding energies for
neutron-deficient Pb isotopes can also be improved by configuration
mixing, as demonstrated in the non-relativistic calculations for the
even-even $^{182-194}\rm Pb$~\cite{Bender2004PR}. For the spherical Ca and Sn isotopes, the energies gained from the restoration of rotational symmetry and configuration mixing are expected to be much smaller as illustrated in the systematic beyond mean-field studies~\cite{Bender2006Phys.Rev.C,Yao2010}. Therefore, the inclusion of these energies will not change significantly the discrepancy between the mean-field results and the corresponding data for such spherical isotopes..

\begin{figure*}[!htbp]
\includegraphics[width=12cm]{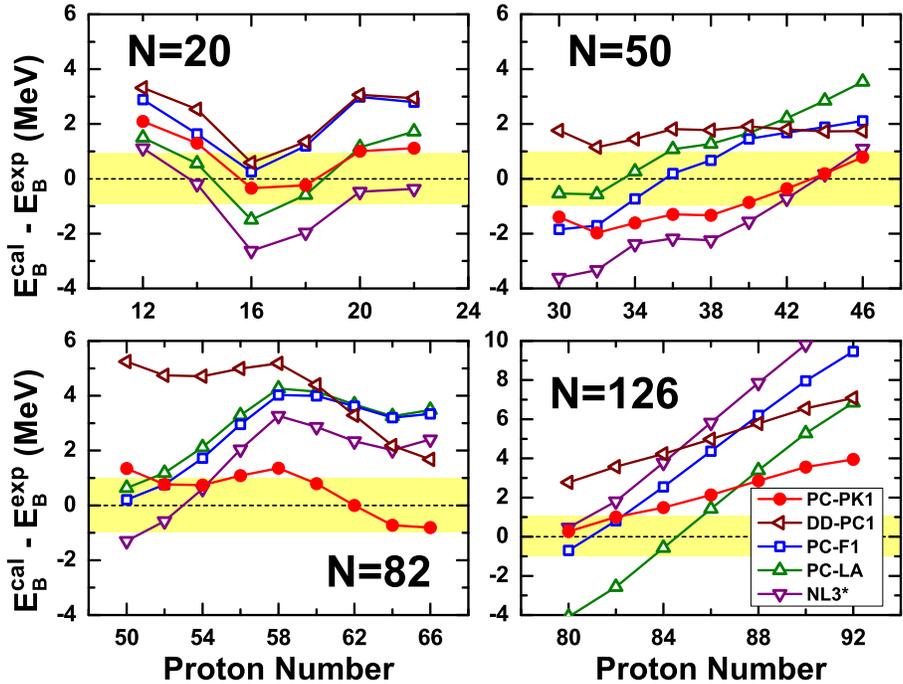}
\caption{(Color online) Same as Fig. 2 but for the N=20, N=50, N=82 and N=126 isotones.}
\label{fig:sph_isotone}
\end{figure*}

The binding energies for the isotonic chains are very important to
examine the balance between the Coulomb field and the isovector
channel of the Lagrangian density in Eq.~(\ref{EQ:LAG}). Here the
binding energies for the $N=20$, $N=50$, $N=82$, and $N=126$ isotones
are calculated with PC-PK1 and their deviations from the
data~\cite{Audi2003NPa} are shown in Fig.~\ref{fig:sph_isotone} in
comparison with those with DD-PC1, PC-F1, PC-LA ,and NL3*. In general, PC-PK1
improves the overall agreement with data in comparison with the
other interactions, especially for $N=82$ and $N=126$ isotones. The
deviations are within $1$ MeV for $N=82$ isotones and $2$ MeV for
both $N=20$ and $N=50$ isotones. A remarkable improvement in the
binding energies as well as proper isospin dependence for $N=126$ is
found in the calculations with PC-PK1. In short, PC-PK1 provides better
prediction for not only the binding energies but also its isospin dependence.

\subsection{Two-neutron separation energy}

\begin{figure}[!htbp]
\includegraphics[width=9cm]{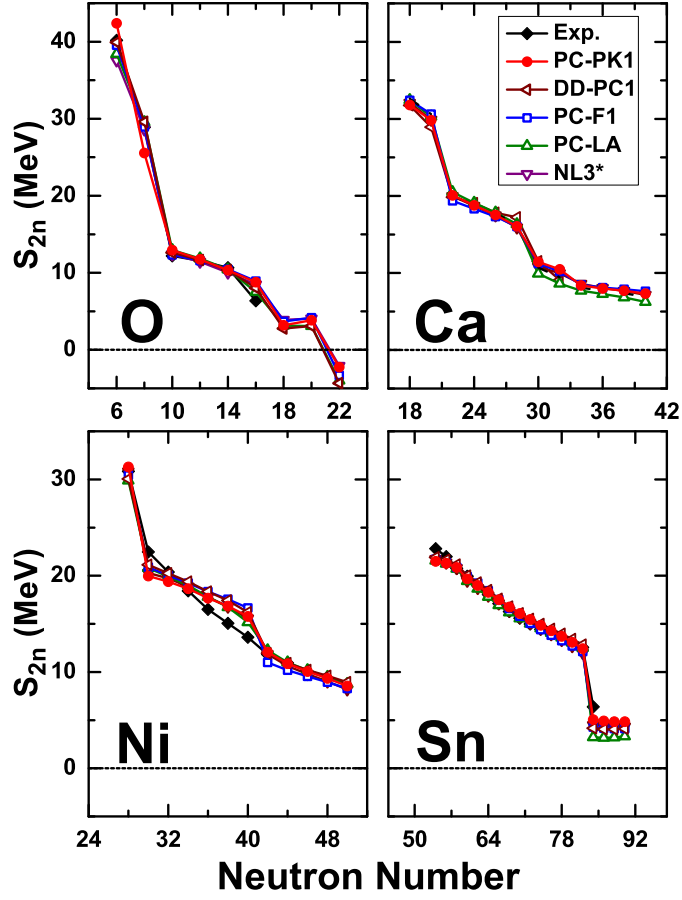}
\caption{(Color online) The calculated two-neutron separation energies for O, Ca, Ni, and Sn isotopes
by PC-PK1 in comparison with data~\cite{Audi2003NPa} and those by DD-PC1, PC-F1, PC-LA, and NL3*.}
\label{fig:sph_S2n}
\end{figure}

From the binding energies, one can extract the two-neutron
separation energies, $S_{2n}=E_B(N,Z)-E_B(N-2,Z)$. In
Fig.~\ref{fig:sph_S2n}, the two-neutron separation energies for
even-even O, Ca, Ni, and Sn isotopes predicted by PC-PK1 are shown
in comparison with data~\cite{Audi2003NPa} and those by DD-PC1, PC-F1, PC-LA,
and NL3*. Generally speaking, similar as the other interactions, the
calculation with PC-PK1 reproduces the experimental two-neutron
separation energies quite well.
For the oxygen isotopic chain, all the effective interactions predict the last bound neutron-rich nucleus as $^{28}\rm O$ contrary to experiment in which $^{24}\rm O$ is so far the last bound neutron-rich nucleus. One can also find that the
deviations are large for the even-even $^{58-68} \rm Ni$ which can
be attributed to the underestimation of binding energies as shown in
Fig.~\ref{fig:sph_isotope}. Moreover, visible deviations between
different predictions can be seen in the neutron-rich Sn isotopes,
which requires future experimental confirmation.

\subsection{Single-particle level}

\begin{figure*}[!htbp]
\includegraphics[width=15cm]{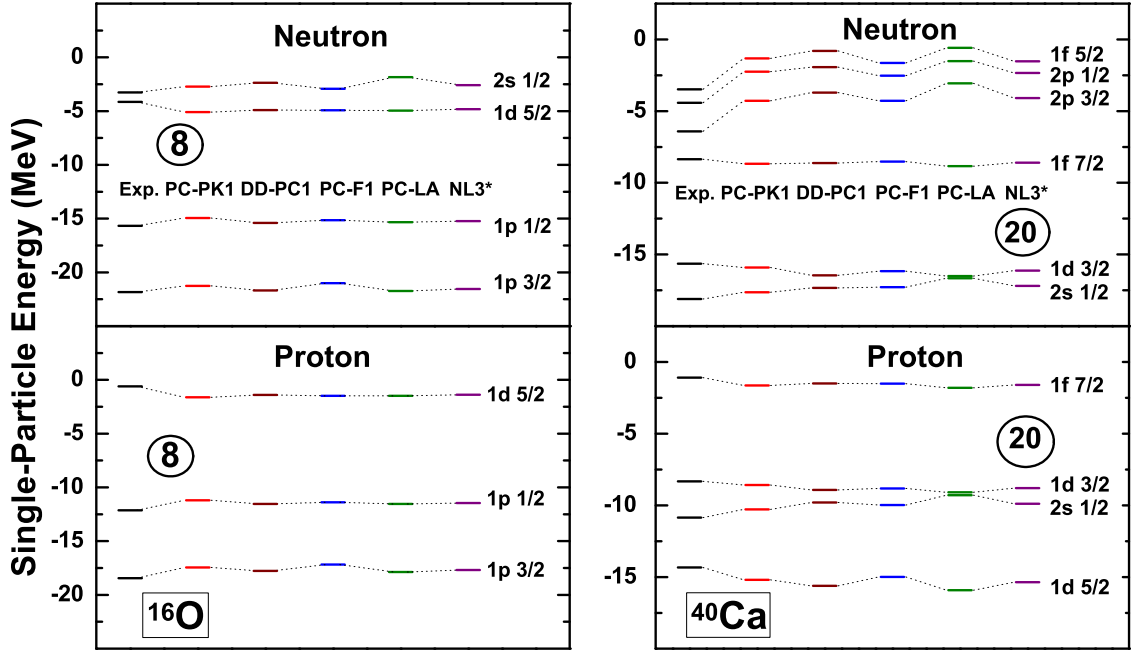}
\caption{(Color online) The calculated single-particle energies for $^{16} \rm O$ and $^{40} \rm Ca$
by PC-PK1 in comparison with data~\cite{Isakov2002EPJ} and those by DD-PC1, PC-F1, PC-LA, and NL3*.}
\label{fig:sph_speOCa}
\end{figure*}

\begin{figure*}[!htbp]
\includegraphics[width=15cm]{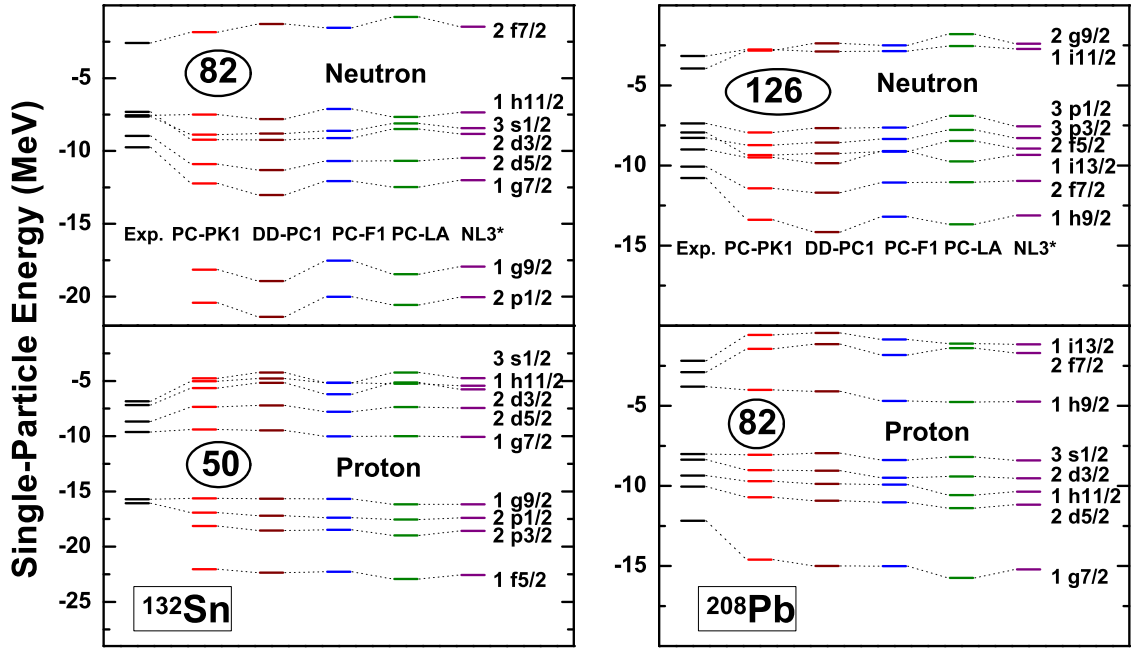}
\caption{(Color online) Same as Fig.~\ref{fig:sph_speOCa} but for $^{132} \rm Sn$ and $^{208} \rm Pb$.}
\label{fig:sph_speSnPb}
\end{figure*}

In Figs.~\ref{fig:sph_speOCa}, \ref{fig:sph_speSnPb}, the
calculated single-particle energies for $^{16} \rm O$, $^{40} \rm
Ca$, $^{132}\rm Sn$, and $^{208}\rm Pb$ by PC-PK1 are shown in
comparison with data~\cite{Isakov2002EPJ} and those by DD-PC1, PC-F1, PC-LA,
and NL3*. The experimental values are extracted from the
single-nucleon separation energies or excitation
energies~\cite{Isakov2002EPJ}. The theoretical single-particle
energies are the eigenvalues of the Dirac equation for nucleon.
It should be kept in mind that the calculations are performed by
neglecting particle-vibration coupling~\cite{Litvinova2006Phys.Rev.C}.

In Figs.~\ref{fig:sph_speOCa}, \ref{fig:sph_speSnPb}, it is clearly
shown that the single-particle levels near the magic numbers and the
corresponding shell gaps given by PC-PK1 are in good agreement with
the experimental values. In particular for $^{16} \rm O$ and $^{40}
\rm Ca$, both the experimental proton and neutron single-particle
spectra are well reproduced by PC-PK1. For $^{132}\rm Sn$ and
$^{208}\rm Pb$, the empirical levels close to the Fermi surface are
also reproduced well. Moreover, the spurious shells at $Z=58$
($^{132}\rm Sn$) and $Z=92$ ($^{208}\rm Pb$) are found for all the
effective interactions, which may be improved by the inclusion of
$\rho$-tensor couplings~\cite{Long2007Phys.Rev.C}.

\subsection{Charge radii and neutron skin thicknesses}

\begin{figure}[!htbp]
\includegraphics[width=8cm]{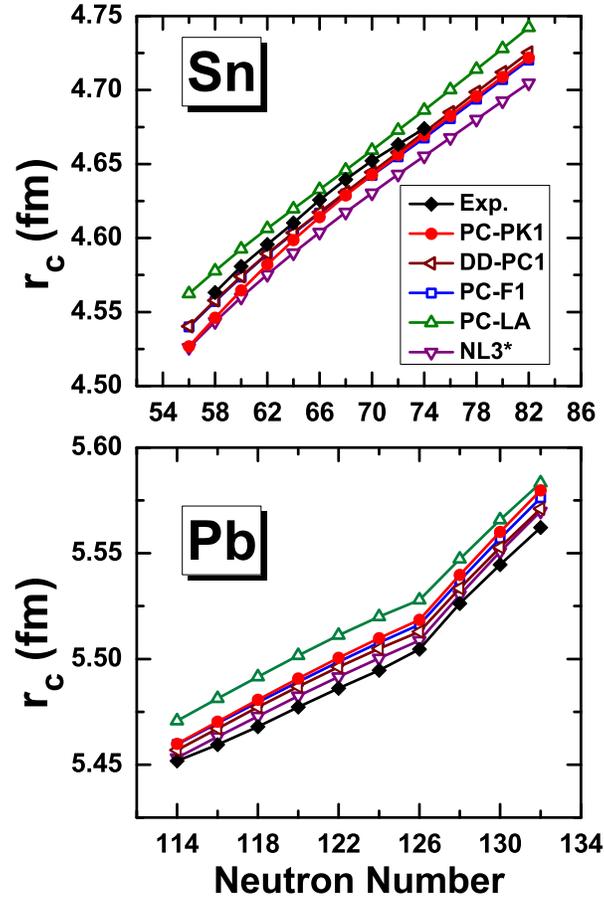}
\caption{(Color online) The calculated charge radii for Sn and Pb isotopes
by the PC-PK1 in comparison with data~\cite{DeVries1987ADNDT,Nadjakov1994ADNDT} and those by
DD-PC1, PC-F1, PC-LA, and NL3*.}
\label{fig:sph_rch}
\end{figure}

In Fig.~\ref{fig:sph_rch}, the charge radii for Sn and Pb isotopes
predicted by PC-PK1 are shown in comparison with
data~\cite{DeVries1987ADNDT,Nadjakov1994ADNDT} and those by DD-PC1, PC-F1,
PC-LA, and NL3*. It is seen that all the effective interactions
reproduce the observed charge radii of Sn isotopes quite well (within
0.3\%). For the Pb isotopes, the kink in the charge radii has been
excellently reproduced by all the effective interactions.
Quantitatively, the observed charge radii of Pb isotopes are reproduced by the calculations with DD-PC1, PC-F1, and NL3* within $\sim 0.3\%$, while the calculations with PC-LA within $\sim 0.5\%$.

\begin{figure}[!htbp]
\includegraphics[width=8cm]{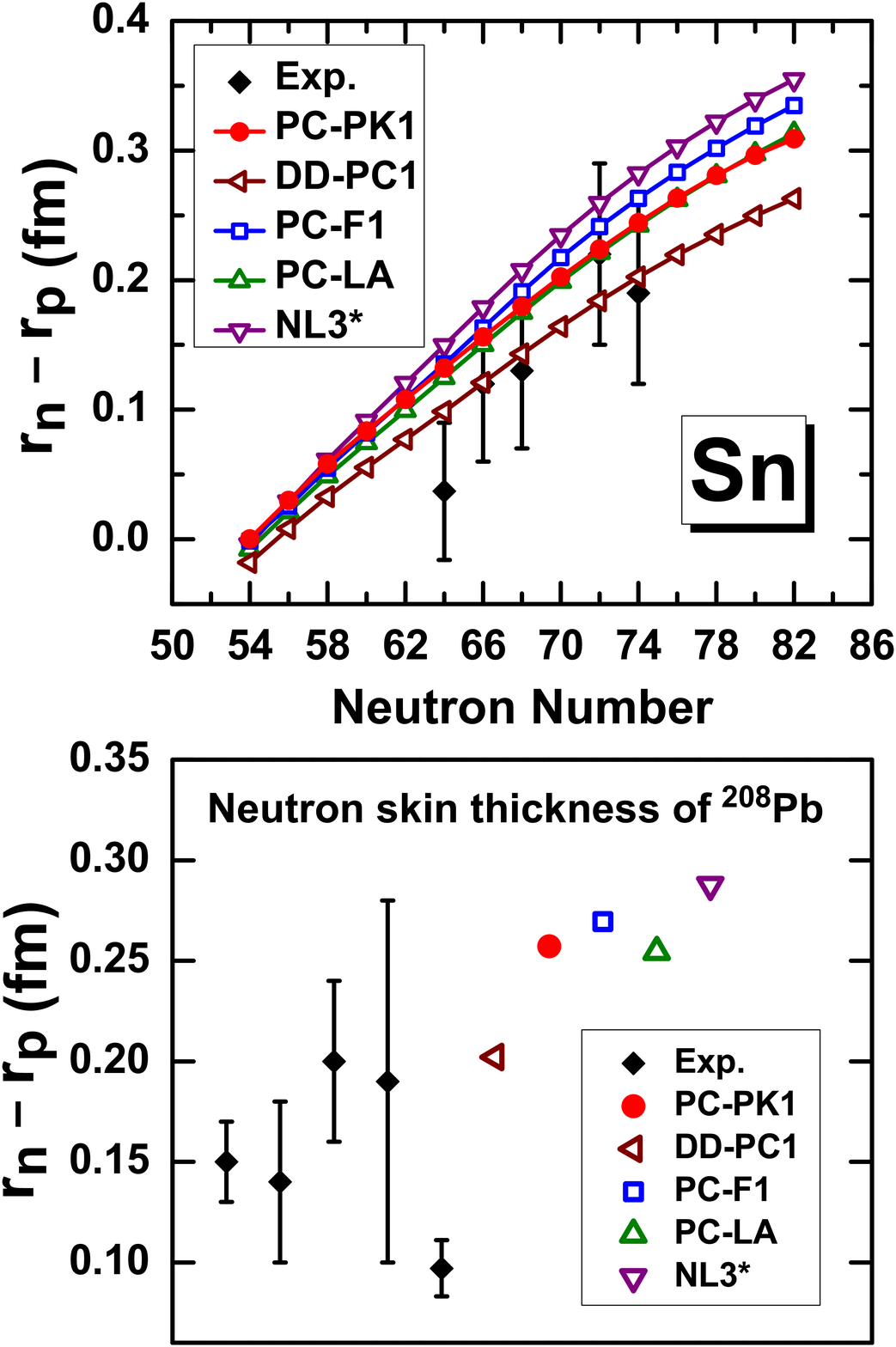}
\caption{(Color online) The calculated neutron skin thicknesses for Sn
isotopes and $^{208}\rm Pb$ by PC-PK1 in comparison with
data~\cite{Krasznahorkay2004NP,Clark2003PR} and those by DD-PC1, PC-F1,
PC-LA, and NL3*. In the lower panel, the data for $^{208}\rm Pb$
deduced from antiprotonic atoms~\cite{Trzcinska2001PRL}, polarized
proton scattering~\cite{Ray1979PR,Hoffmann1981PRL}, elastic proton
scattering~\cite{Starodubsky1994PR}, inelastic $\alpha$
scattering~\cite{Krasznahorkay1994NP}, and proton-nucleus elastic
scattering~\cite{Clark2003PR} are shown from left to right
respectively.} \label{fig:sph_spenth}
\end{figure}

In Fig.~\ref{fig:sph_spenth}, the neutron skin thicknesses for Sn
isotopes and $^{208}\rm Pb$ predicted by PC-PK1 are shown in
comparison with data~\cite{Krasznahorkay2004NP,Clark2003PR} and
those by DD-PC1, PC-F1, PC-LA, and NL3*. For Sn isotopes, although
PC-PK1 slightly overestimates the neutron skin thickness in
comparison with DD-PC1, it nicely reproduces the isotopic trend. For
$^{208}\rm Pb$, all the interactions except DD-PC1 give similar
neutron skin thicknesses which are larger than the data deduced from
antiprotonic atoms~\cite{Trzcinska2001PRL}, polarized proton
scattering~\cite{Ray1979PR,Hoffmann1981PRL}, elastic proton
scattering~\cite{Starodubsky1994PR}, proton-nucleus elastic
scattering~\cite{Clark2003PR}, and agree within the experimental
error bar with that from inelastic $\alpha$
scattering~\cite{Krasznahorkay1994NP}. The slightly overestimated
neutron skin thicknesses are due to the enhanced symmetry energies
for nuclear matter shown in Table~\ref{Tabel:nuclearmatter}. In overall, the DD-PC1 parametrization provides better description of experimental charge radii and neutron skin thickness due to its smaller symmetry energy at saturation density.

\section{Deformed nuclei}

In this section, we will focus on the description of the binding
energies and deformations for selected well-deformed even-even
nuclei. In order to investigate the fission barrier, a constrained
calculation is also carried out by taking $^{240} \rm Pu$ as an
example.

\subsection{Binding energy and deformation}

The binding energies and quadrupole deformations of the ground
states for Yb and U isotopes are investigated in axially deformed
code with PC-PK1 in comparison with those with DD-PC1 and PC-F1.

\begin{figure*}[!htbp]
\includegraphics[width=12cm]{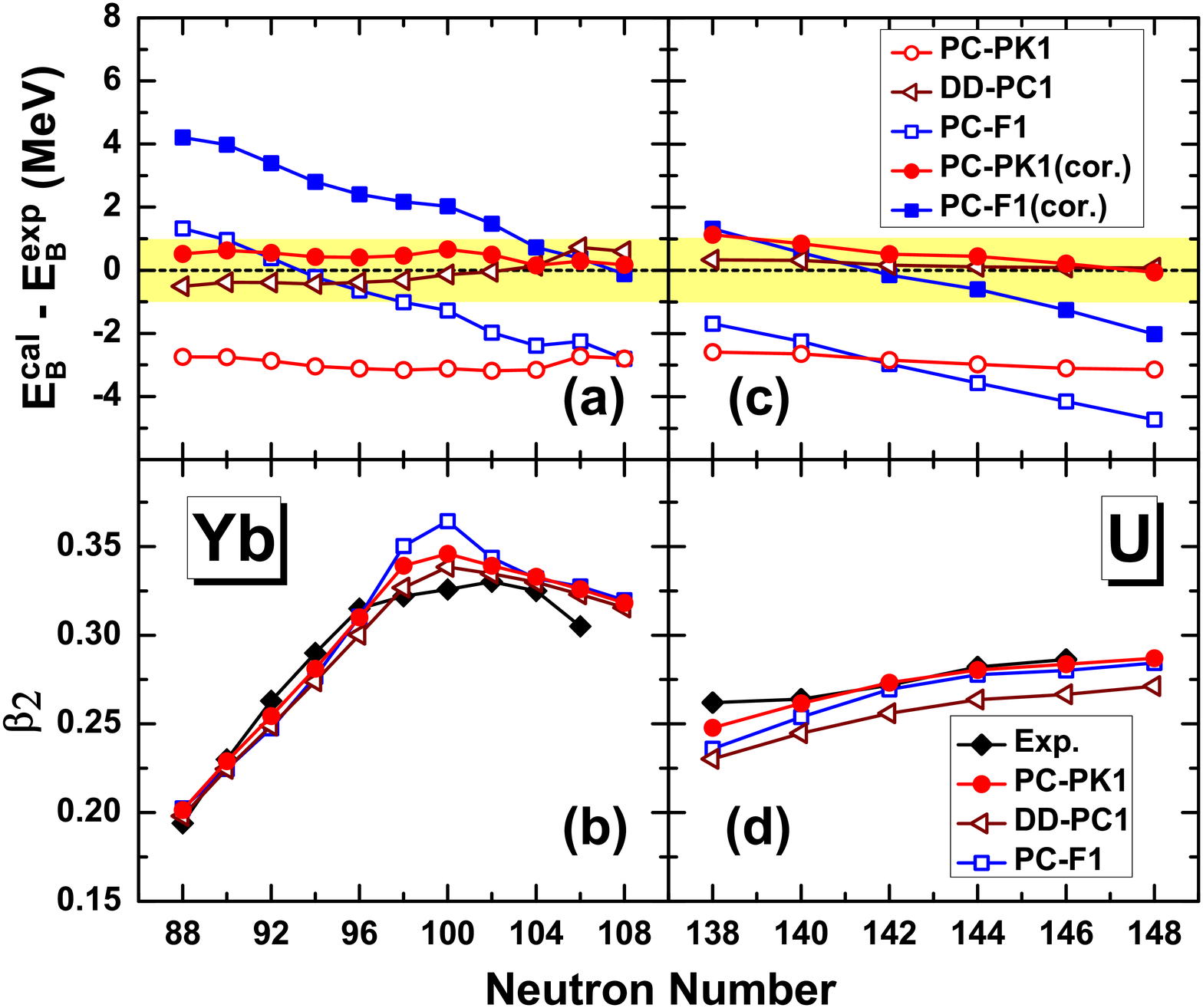}
\caption{(Color online) Deviations of the calculated binding
energies from the data~\cite{Audi2003NPa} for Yb and U isotopes in
axially deformed code by PC-PK1, DD-PC1 and PC-F1 (upper panel) as well as
the corresponding calculated ground-state deformations in comparison
with data~\cite{Raman2001ADNDT} (lower panel). The filled circles
and squares in the upper panels correspond to the rotational
corrected ones given by PC-PK1 and PC-F1 respectively.}
\label{fig:def_YbU}
\end{figure*}

In the upper panels of Fig.~\ref{fig:def_YbU}, the deviations of the
calculated binding energies with PC-PK1, DD-PC1, and PC-F1 from the
data~\cite{Audi2003NPa} are shown as circles, triangles, and squares
respectively.

Before taking into account the rotational correction for the binding
energies, a systematical underestimate of the binding energies
around $3$ MeV for both Yb and U isotopes is found for PC-PK1. For
PC-F1, the difference between the calculated and the observed
binding energy decreases monotonically with the isospin values,
i.e., around $1 \sim -3$ MeV for Yb isotopes and $-2 \sim -5$ MeV
for U isotopes. As almost all the isotopes shown in
Fig~\ref{fig:def_YbU} are used to adjust the parameters, the
predicted binding energies by DD-PC1 are in good agreement with the
data (within $1$ MeV).

After taking into account the energy correction due to the
restoration of rotational symmetry in the cranking
approximation~\cite{Girod1979NP}, the calculated results by PC-PK1
(filled circles) reproduces the data quite well for both Yb and U
isotopes, and the deviations are within $1$ MeV. While the
differences between the corrected binding energies given by PC-F1
(filled squares) and data are still large. Since DD-PC1 is adjusted
to the binding energies of 64 well-deformed nuclei, the rotational
correction energy is not considered in the corresponding
calculations.
The energy correction due to the restoration of rotational symmetry can be taken into account with the microscopic treatment of angular momentum projection or the cranking approximation~\cite{Ring1980B}. It is noted that difference between the cranking approximation and the angular momentum projection exists (for example, it reaches $1$ MeV in $^{240}$Pu)~\cite{Bender2004Phys.Rev.C}. For simplicity of systematic calculations, only the cranking approximation is used here.

In the lower panels of Fig.~\ref{fig:def_YbU}, the calculated quadrupole
deformations for the ground states by PC-PK1, DD-PC1, and PC-F1 are given in
comparison with the corresponding data~\cite{Raman2001ADNDT}. It
shows that the deformations and their corresponding evolutions with
neutron number for both Yb and U isotopes are well reproduced by
PC-PK1, DD-PC1, and PC-F1.

\subsection{Fission barrier}

\begin{figure}[!htbp]
\includegraphics[width=8cm]{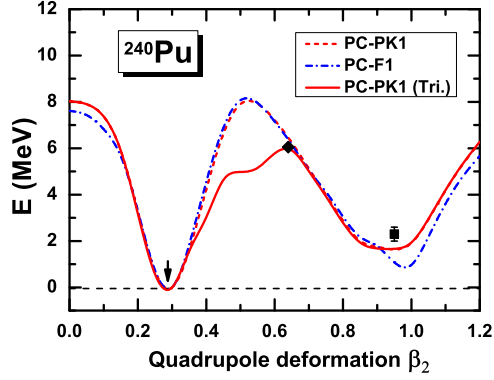}
\caption{(Color online) The potential energy curves for $^{240} \rm
Pu$ as functions of the quadrupole deformation $\beta_2$ in the
calculations with PC-PK1. The dashed and solid lines correspond to
the axial results and the triaxial results, i.e., the minima for each $\beta$ for the
potential energy surface (PES) in the $\beta-\gamma$ plane, respectively.
For comparison, the
axially-symmetric result given by PC-F1 is also included as
dot-dashed line. The data for the ground-state
deformation~\cite{Raman2001ADNDT}, the barrier height~\cite{RIPL-2},
and the energy of the fission isomer \cite{Bjornholm1980RMP} are
respectively indicated by an arrow, a diamond, and a square. To
guide the eyes, the diamond and square are respectively set at
$\beta_2=0.64$ and $\beta_2=0.95$.} \label{fig:def_Pu}
\end{figure}

In Fig.~\ref{fig:def_Pu}, the potential energy curves for $^{240}
\rm Pu$ as functions of the quadrupole deformation $\beta_2$ are
shown. The dashed and solid lines correspond to the
axially-symmetric and the triaxial calculations with PC-PK1, respectively.
In the case of triaxial calculation, the solid line refers to the minima for each $\beta$ for the potential energy surface (PES) in the $\beta-\gamma$ plane.
For comparison, the axially-symmetric result given by PC-F1 is also included.

It is found that the PC-PK1 provides not only a good description for
the deformation of the ground state~\cite{Raman2001ADNDT} but also
the energy difference between the ground-state and the shape
isomeric state~\cite{Bjornholm1980RMP}. Furthermore, after including
the triaxiality, as shown in Fig~\ref{fig:def_Pu}, the fission
barrier given by PC-PK1 is in agreement with the empirical
value~\cite{RIPL-2}. It should be noted that the pairing correlation plays an important role in the description of fission barrier. Discussion on the dependence of the fission barrier height on the pairing correlations can be found in Ref.~\cite{Karatzikos2010PL}.

\section{Nuclear excited properties}

As a test of the new parameter set PC-PK1 in the description of
nuclear spectroscopic properties for low-lying excitation states,
the collective excitation spectra and transition probabilities in
$^{150} \rm Nd$ as well as the characteristic collective observables
for Nd isotopes will be calculated starting from a five-dimensional
collective Hamiltonian in which the parameters are determined by
constrained self-consistent RMF calculations for triaxial
shapes~\cite{Niksic2009PR,Li2009PR,Li2009PRa}.

\begin{figure}[!htbp]
\includegraphics[width=8cm]{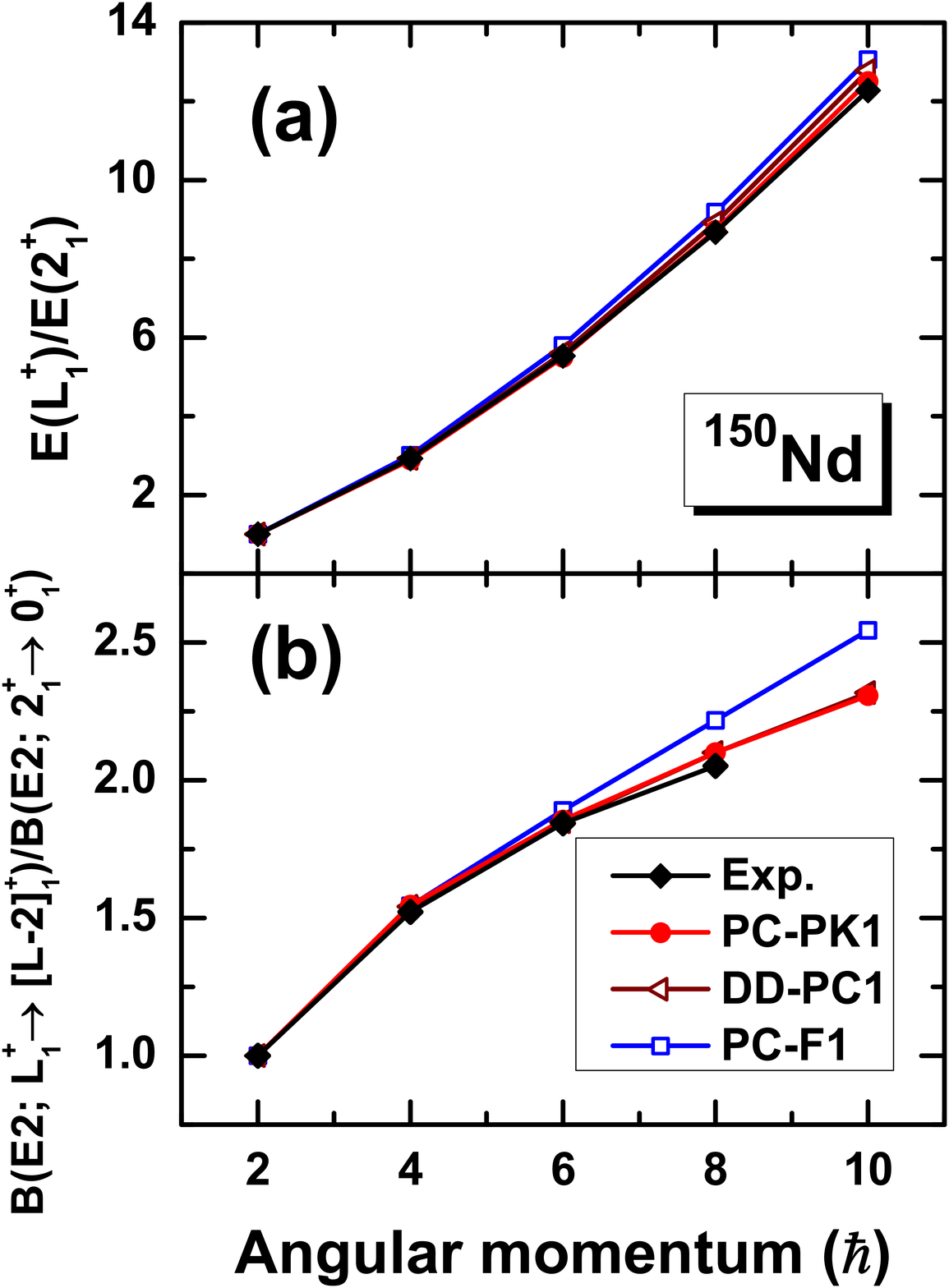}
\caption{(Color online) The predicted excitation energies (upper
panel) and $B(E2;L^+_1\rightarrow [L-2]^+_1)$ values (lower panel)
for the yrast states in $^{150} \rm Nd$ by PC-PK1 in comparison with
data~\cite{NNDC,LBNL} and those by DD-PC1 and PC-F1. The energies
and $B(E2)$ values are respectively normalized to $E(2^+_{1})$ and
$B(E2;2^+_1\rightarrow 0^+_1)$.} \label{fig:Exc_E2}
\end{figure}

In Fig.~\ref{fig:Exc_E2}, the excitation energies and
$B(E2;L^+_1\rightarrow [L-2]^+_1)$ values for the yrast states in
$^{150} \rm Nd$ predicted by PC-PK1 are shown in comparison with
data~\cite{NNDC,LBNL} and those by DD-PC1 and PC-F1. It can be seen
that all the effective interactions provide similar excitation
energies and intraband $B(E2)$ values for the yrast band and
reproduce the data quite well.

\begin{figure}[!htbp]
\includegraphics[width=10cm]{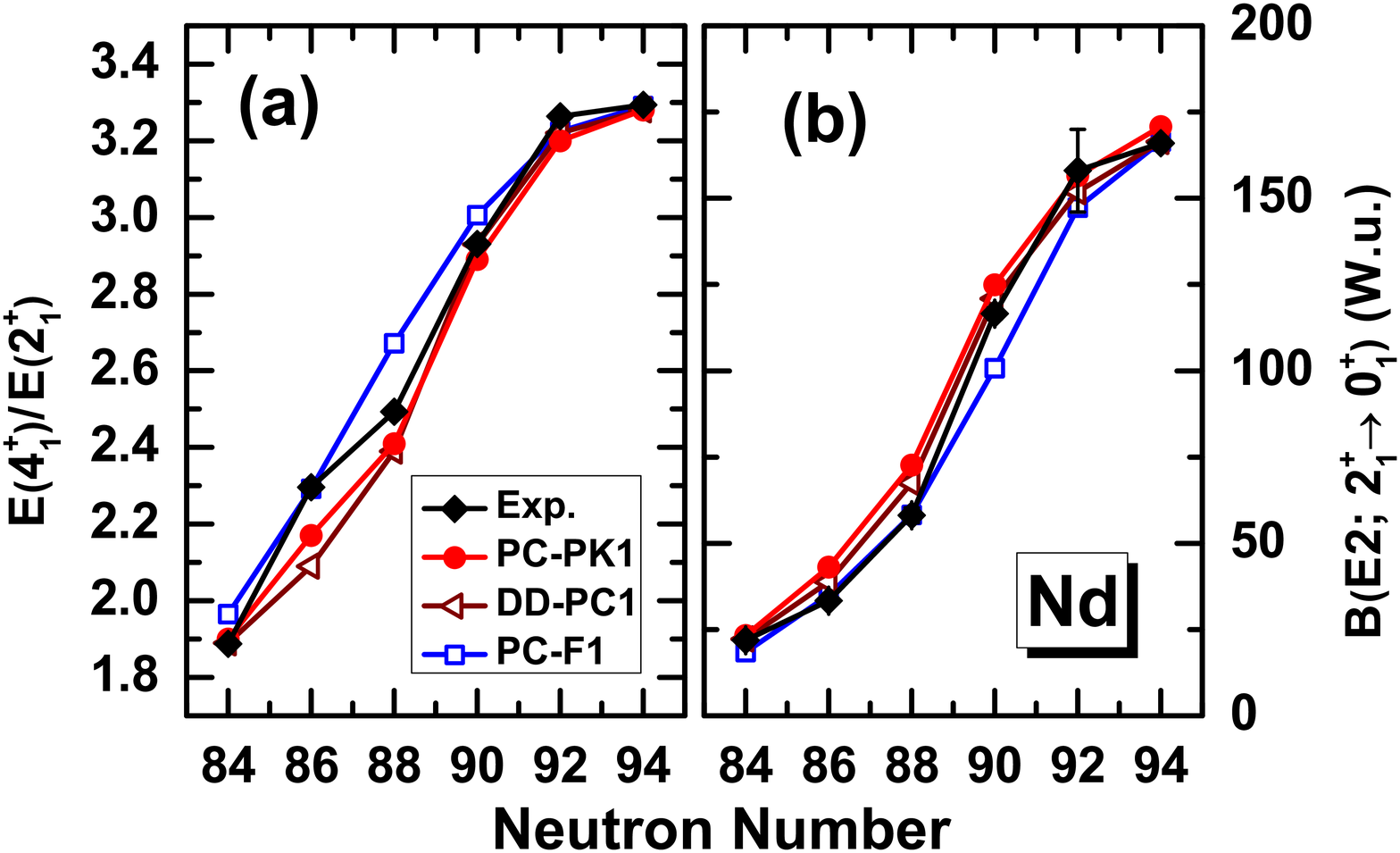}
\caption{(Color online) The predicted characteristic collective observables $R_{4/2}=E(4^+_1)/E(2^+_1)$
and $B(E2;2^+_1\rightarrow 0^+_1)$ (in W.u.) for Nd isotopes by PC-PK1 in comparison with
data~\cite{NNDC,LBNL} and those by DD-PC1 and PC-F1.}
\label{fig:Exc_R4}
\end{figure}

In Fig.~\ref{fig:Exc_R4}, the characteristic collective observables
$R_{4/2}=E(4^+_1)/E(2^+_1)$ and $B(E2;2^+_1\rightarrow 0^+_1)$ for
Nd isotopes given by PC-PK1 are shown in comparison with
data~\cite{NNDC,LBNL} and those by DD-PC1 and PC-F1. It is found
that all the parameter sets reproduce the data quite well. In
particular, the calculations reproduce in detail the rapid increase
of $R_{4/2}$ and $B(E2)$ with the neutron number, i.e., from
$R_{4/2}\sim1.9$ and $B(E2)<30$ W.u. in near spherical $^{144}\rm
Nd$ to $R_{4/2}\sim3.3$ and $B(E2)>150$ W.u. in well-deformed
$^{152}\rm Nd$.

It shows clearly that the new effective interaction PC-PK1 can
provide a good description not only for the ground state properties
in spherical and deformed nuclei but also for the nuclear spectroscopic
properties of low-lying excitation states.

\section{Summary}

In summary, a new parametrization PC-PK1 for the nuclear covariant
energy density functional with nonlinear point-coupling interaction
has been proposed by fitting to observables of 60 selected spherical
nuclei, including the binding energies, charge radii and empirical
pairing gaps. By scaling the coupling constants in PC-PK1 in
accordance with the QCD-based Lagrangian, it is found that all the
nine parameters are natural. The success of PC-PK1 has been
illustrated through the description for infinite nuclear matter and
finite nuclei including the ground-state and low-lying excited
states.

For the spherical nuclei, PC-PK1 can provide better descriptions for
the binding energies in comparison with DD-PC1, PC-F1, PC-LA, and
NL3* sets. For neutron skin thicknesses, the DD-PC1 provides better description as compared with the other effective interactions due to its smaller symmetry energy at saturation density.

Taking Yb and U isotopes as examples, it is found that the PC-PK1
reproduces the deformations and their corresponding evolutions with
neutron number quite well. After taking into account the rotational correction energy in the cranking approximation, the binding energies given by PC-PK1 are in very good agreement with data within 1 MeV, which indicates that PC-PK1 achieves the same quality as DD-PC1 in the description for deformed nuclei. Moreover, PC-PK1 provides good description for isospin dependence of binding energy along either the isotopic or the isotonic chains, which makes it reliable for application in exotic nuclei. It is noted that the rotational correction energy evaluated using the cranking approximation may differ from that using angular momentum projection.

Constrained calculations have also been performed for $^{240}\rm Pu$
in order to investigate the fission barrier. It is found that the
PC-PK1 provides not only a good description for the deformation of
the ground state~\cite{Raman2001ADNDT} but also the energy
difference between the ground-state and the shape isomeric
state~\cite{Bjornholm1980RMP}. Furthermore, after including the
triaxiality, the fission barrier given by PC-PK1 is in agreement with the empirical value~\cite{RIPL-2}.

The predictive power of the PC-PK1 is also illustrated in the
description for the collective excitation spectra and transition
probabilities in $^{150} \rm Nd$ as well as the characteristic
collective observables for Nd isotopes in a five-dimensional
collective Hamiltonian in which the parameters are determined by
constrained calculations for triaxial shapes. There are also many
extensions of nuclear covariant energy density functional theory
beyond mean-field using projection techniques~\cite{Yao2009PR} and
generator coordinate methods~\cite{Niksic2006PR,Yao2010Phys.Rev.C}.
More microscopic analysis of nuclear low-lying states in context of
these frameworks with PC-PK1 is in progress.

The density-dependent parametrization DD-PC1 is determined mainly from the masses of deformed nuclei and the EOS of nuclear matter. However, the calculations of the rearrangement terms for the density-dependent parametrization can be nontrivial in some cases, in particular for RPA calculations. Here the nonlinear parametrization PC-PK1 has been optimized to the masses, charge radii and empirical pairing gaps for selected 60 spherical nuclei. It has been illustrated that the PC-PK1 can provide very good descriptions for both spherical and deformed nuclei. Therefore, the non-linear parametrization is very useful as it combines the simplicity with very good predictions for many nuclear properties.

\begin{acknowledgments}
We thank P. Ring and T. Nik\v{s}i\'{c} for stimulating discussions
and kind help in the comparison with density-dependent
point-coupling results. This work was partly supported by the Major
State 973 Program 2007CB815000 and the NSFC under Grant Nos.
10775004, 10947013 and 10975008 and the Southwest University Initial
Research Foundation Grant to Doctor (No. SWU109011).
\end{acknowledgments}

\newpage %Just because of unusual number of tables stacked at end

\newpage
\begin{table}[!htbp]
\caption{The point-coupling constants and pairing strengths of PC-PK1 set.
 The corresponding QCD-scaled coupling constants $c_{ln}$ are given in the last column as well.}
\label{Tabel:PC-PK1}
\begin{ruledtabular}
\begin{tabular}{ccccc}
Coupling Constant & Value                    &  Dimension          & $c_{ln}$\\
\hline
 $~\alpha_S$      &$-3.96291\times10^{-4}$   &  ${\rm MeV}^{-2}$   &-1.695 \\

 $~\beta_S$       &$8.6653\times10^{-11}$    &  ${\rm MeV}^{-5}$   &1.628 \\

 $~\gamma_S$      &$-3.80724\times10^{-17}$  &  ${\rm MeV}^{-8}$   &-3.535 \\

 $~\delta_S$      &$-1.09108\times10^{-10}$  &  ${\rm MeV}^{-4}$   &-0.277 \\

 $~\alpha_V$      &$2.6904\times10^{-4}$     &  ${\rm MeV}^{-2}$   &1.151 \\

 $~\gamma_V$      &$-3.64219\times10^{-18}$  &  ${\rm MeV}^{-8}$   &-0.338\\

 $~\delta_V$      &$-4.32619\times10^{-10}$  &  ${\rm MeV}^{-4}$   &-1.097\\

$~\alpha_{TV}$    &$2.95018\times10^{-5}$    &  ${\rm MeV}^{-2}$   &0.505\\

$~\delta_{TV}$    &$-4.11112\times10^{-10}$  &  ${\rm MeV}^{-4}$   &-4.171\\
\hline
      $~V_n$      &$-349.5$                  &  ${\rm MeV~fm}^3$   &~    \\

      $~V_p$      &$-330$                    &  ${\rm MeV~fm}^3$   &~     \\
\end{tabular}
\end{ruledtabular}
\end{table}

\setlength{\LTcapwidth}{5.9in}
\begin{longtable*}{@{\extracolsep{0.3in}}ccccccc}
\caption{The calculated binding energies (in MeV) for selected spherical
nuclei by PC-PK1 in comparison with the data~\cite{Audi2003NPa} and
those by DD-PC1~\cite{Niksic2008PR}, PC-F1~\cite{Burvenich2002PR}, PC-LA \cite{Nikolaus1992PR},
and NL3*~\cite{Lalazissis2009PL}. The bold-faced quantities denote
that the experimental values of the corresponding nuclei are used in
the parametrization fitting. The root mean square (rms) deviation
$\displaystyle\Delta=\sum\limits_i^N\sqrt{(E_i^{\rm exp}-E_i^{\rm
cal})^2/N}$ and the root of relative square (rrs) deviation
$\displaystyle\delta=\sum\limits_i^N\sqrt{\frac{(E_i^{\rm
exp}-E_i^{\rm cal})^2/(E_i^{\rm exp})^2}{N}}$ are respectively
listed in the last two rows.}
\label{Tabel:mass}\\
\hline\hline
\noalign{\vspace{3pt}}
        Nuclei  &       Exp. &     PC-PK1 &     DD-PC1 &    PC-F1 &      PC-LA &        NL3* \\
\noalign{\vspace{3pt}}
\hline
\noalign{\vspace{3pt}}
\endfirsthead
\caption{(Continued).}\\%
\hline\hline
\noalign{\vspace{3pt}}
        Nuclei  &       Exp. &     PC-PK1 &     DD-PC1 &    PC-F1 &      PC-LA &        NL3* \\
\noalign{\vspace{3pt}}
\hline
\noalign{\vspace{3pt}}
\endhead
\hline
\endfoot \hline\hline
\endlastfoot

        $^{16}\rm O$ &  127.619  &  \textbf{127.280}  &   128.527   &  \textbf{127.691} &   \textbf{127.407}& \textbf{128.112 } \\
        $^{18}\rm O$ &  139.806  &  \textbf{140.223}  &   141.145   &          140.028  &           140.356 &         140.504   \\
        $^{20}\rm O$ &  151.370  &  \textbf{151.962}  &   152.790   &          151.606  &           152.228 &         151.955   \\
        $^{22}\rm O$ &  162.026  &  \textbf{162.285}  &   163.141   &          162.054  &           162.665 &         161.990   \\
       $^{18}\rm Ne$ &  132.143  &  \textbf{132.088}  &   132.923   &          132.216  &           132.317 &         132.494   \\
       $^{20}\rm Mg$ &  134.468  &  \textbf{134.563}  &   135.141   &          134.613  &           134.992 &         134.786   \\
       $^{34}\rm Si$ &  283.429  &  \textbf{284.727}  &   285.967   &          285.067  &           283.989 &         283.236   \\
        $^{36}\rm S$ &  308.714  &  \textbf{308.374}  &   309.305   &          308.973  &           307.221 &         306.086   \\
       $^{38}\rm Ar$ &  327.342  &  \textbf{327.107}  &   328.691   &          328.540  &           326.755 &         325.379   \\
       $^{36}\rm Ca$ &  281.360  &  \textbf{281.412}  &   281.878   &          282.001  &           280.454 &         279.579   \\
       $^{38}\rm Ca$ &  313.122  &  \textbf{313.230}  &   314.501   &          314.415  &           312.901 &         311.669   \\
       $^{40}\rm Ca$ &  342.052  &  \textbf{343.060}  &   345.113   &  \textbf{345.041} &           343.202 & \textbf{341.578 } \\
       $^{42}\rm Ca$ &  361.896  &  \textbf{363.142}  &   365.143   &          364.411  &           363.685 &         361.547   \\
       $^{44}\rm Ca$ &  380.960  &  \textbf{381.915}  &   383.967   &          382.748  &           382.789 &         380.246   \\
       $^{46}\rm Ca$ &  398.769  &  \textbf{399.451}  &   401.668   &          400.060  &           400.627 &         397.718   \\
       $^{48}\rm Ca$ &  415.990  &  \textbf{415.492}  &   417.973   &  \textbf{416.085} &           416.969 & \textbf{413.616 } \\
       $^{50}\rm Ca$ &  427.490  &  \textbf{426.937}  &   428.660   &          427.302  &           426.883 &         424.445   \\
       $^{42}\rm Ti$ &  346.905  &  \textbf{348.024}  &   349.848   &          349.701  &           348.626 &         346.539   \\
       $^{50}\rm Ti$ &  437.781  &  \textbf{436.445}  &   437.761   &          436.171  &           437.223 &         434.389   \\
       $^{56}\rm Ni$ &  483.992  &  \textbf{483.669}  &   481.447   &  \textbf{480.758} &           481.826 &         481.058   \\
       $^{58}\rm Ni$ &  506.458  &         {503.636}  &   502.587   &  \textbf{501.646} &           502.623 &         501.342   \\
       $^{72}\rm Ni$ &  613.169  &  \textbf{614.875}  &   617.071   &          614.646  &           614.486 & \textbf{612.561}   \\
       $^{84}\rm Se$ &  727.343  &  \textbf{725.732}  &   728.792   &          726.609  &           727.605 &         724.965   \\
       $^{86}\rm Kr$ &  749.234  &  \textbf{747.939}  &   751.050   &          749.427  &           750.313 &         747.055   \\
       $^{88}\rm Sr$ &  768.468  &  \textbf{767.138}  &   770.240   &  \textbf{769.143} &   \textbf{769.742}&         766.225   \\
       $^{90}\rm Zr$ &  783.892  &  \textbf{783.033}  &   785.806   &  \textbf{785.348} &           785.565 & \textbf{782.336 } \\
       $^{92}\rm Mo$ &  796.508  &  \textbf{796.148}  &   798.308   &          798.191  &           798.719 &         795.788   \\
       $^{94}\rm Ru$ &  806.848  &  \textbf{807.034}  &   808.575   &          808.731  &           809.695 &         807.019   \\
       $^{98}\rm Cd$ &  821.067  &  \textbf{822.765}  &   823.162   &          823.668  &           825.580 &         823.347   \\
      $^{100}\rm Sn$ &  824.794  &         {827.715}  &   827.609   &  \textbf{828.156} &           830.582 &         828.529   \\
      $^{106}\rm Sn$ &  893.868  &  \textbf{892.323}  &   893.469   &          893.370  &           895.447 &         893.873   \\
      $^{108}\rm Sn$ &  914.626  &  \textbf{913.179}  &   914.627   &          914.236  &           916.165 &         914.665   \\
      $^{112}\rm Sn$ &  953.532  &         {951.831}  &   953.922   &  \textbf{953.367} &           954.258 &         952.866   \\
      $^{116}\rm Sn$ &  988.684  &         {987.601}  &   990.019   &         {989.326} &           989.016 & \textbf{987.920 } \\
      $^{120}\rm Sn$ & 1020.546  &        {1020.415}  &   1022.902  & \textbf{1021.704} &          1020.767 &        1020.014   \\
      $^{122}\rm Sn$ & 1035.529  & \textbf{1035.860}  &   1038.417  &         1036.755  &          1035.794 &        1035.116   \\
      $^{124}\rm Sn$ & 1049.963  & \textbf{1050.715}  &   1053.402  & \textbf{1051.160} &          1050.327 &\textbf{1049.631 } \\
      $^{126}\rm Sn$ & 1063.889  & \textbf{1064.993}  &   1067.877  &         1064.978  &          1064.381 &        1063.560   \\
      $^{128}\rm Sn$ & 1077.346  & \textbf{1078.688}  &   1081.835  &         1078.234  &          1077.945 &        1076.885   \\
      $^{130}\rm Sn$ & 1090.293  & \textbf{1091.774}  &   1095.253  &         1090.930  &          1090.993 &        1089.566   \\
      $^{132}\rm Sn$ & 1102.851  & \textbf{1104.202}  &   1108.096  & \textbf{1103.057} &          1103.484 &\textbf{1101.551 } \\
      $^{134}\rm Sn$ & 1109.235  & \textbf{1109.253}  &   1112.253  &         1107.330  &          1106.707 &        1106.027   \\
      $^{134}\rm Te$ & 1123.434  & \textbf{1124.205}  &   1128.176  &         1124.193  &          1124.613 &        1122.859   \\
      $^{136}\rm Xe$ & 1141.878  & \textbf{1142.621}  &   1146.587  & \textbf{1143.601} &          1143.997 &        1142.480   \\
      $^{138}\rm Ba$ & 1158.292  & \textbf{1159.381}  &   1163.283  &         1161.245  &          1161.575 &        1160.331   \\
      $^{140}\rm Ce$ & 1172.692  & \textbf{1174.054}  &   1177.868  &         1176.722  &          1176.953 &        1175.954   \\
      $^{142}\rm Nd$ & 1185.141  & \textbf{1185.938}  &   1189.537  &         1189.138  &          1189.292 &        1188.002   \\
      $^{144}\rm Sm$ & 1195.736  & \textbf{1195.736}  &   1199.024  & \textbf{1199.353} &          1199.420 &        1198.079   \\
      $^{146}\rm Gd$ & 1204.435  & \textbf{1203.712}  &   1206.614  &         1207.635  &          1207.687 &        1206.449   \\
      $^{148}\rm Dy$ & 1210.780  & \textbf{1209.974}  &   1212.454  &         1214.117  &          1214.258 &        1213.186   \\
      $^{150}\rm Er$ & 1215.331  & \textbf{1214.624}  &   1216.686  &         1218.943  &          1219.236 &        1218.343   \\
      $^{206}\rm Hg$ & 1621.049  & \textbf{1621.321}  &   1623.820  &         1620.353  &          1616.956 &        1621.515   \\
      $^{200}\rm Pb$ & 1576.354  & \textbf{1574.885}  &   1577.817  &         1575.666  &          1575.769 &        1578.189   \\
      $^{202}\rm Pb$ & 1592.187  & \textbf{1591.172}  &   1594.139  &         1591.675  &          1591.240 &        1593.909   \\
      $^{204}\rm Pb$ & 1607.506  & \textbf{1607.068}  &   1610.026  &         1607.325  &          1606.187 &\textbf{1609.199}   \\
      $^{206}\rm Pb$ & 1622.324  & \textbf{1622.525}  &   1625.385  &         1622.563  &          1620.490 &        1624.008   \\
      $^{208}\rm Pb$ & 1636.430  & \textbf{1637.438}  &   1640.008  & \textbf{1637.241} & \textbf{1633.865} &\textbf{1638.237}  \\
      $^{210}\rm Pb$ & 1645.552  & \textbf{1645.449}  &   1648.272  &         1644.793  &          1641.484 &        1645.954   \\
      $^{212}\rm Pb$ & 1654.514  & \textbf{1653.425}  &   1656.428  &         1652.275  &          1648.887 &        1653.546   \\
      $^{214}\rm Pb$ & 1663.291  & \textbf{1661.397}  &   1664.481  & \textbf{1659.697} &          1656.073 &\textbf{1661.056 } \\
      $^{210}\rm Po$ & 1645.212  & \textbf{1646.703}  &   1649.441  &         1647.760  &          1644.643 &\textbf{1648.995 }  \\
      $^{212}\rm Rn$ & 1652.497  & \textbf{1654.632}  &   1657.476  &         1656.863  &          1653.921 &        1658.319   \\
      $^{214}\rm Ra$ & 1658.315  & \textbf{1661.172}  &   1664.092  &         1664.512  &          1661.709 &        1666.174   \\
      $^{216}\rm Th$ & 1662.689  & \textbf{1666.248}  &   1669.244  &         1670.649  &          1667.967 &        1672.505   \\
       $^{218}\rm U$ & 1665.648  & \textbf{1669.602}  &   1672.733  &         1675.109  &          1672.491 &        1677.091   \\
\hline
\noalign{\vspace{5pt}}
         $\Delta$    &           &              1.33  &     3.09    &             2.60  &             2.64  &             2.88 \\
         $\delta$    &           &             0.18\% &     0.45\%  &            0.32\% &            0.30\% &            0.34\% \\
\end{longtable*}

\begin{table}[!htbp]
\caption{The calculated charge radii (in fm) for selected spherical
 nuclei by PC-PK1 in comparison with data~\cite{DeVries1987ADNDT,Nadjakov1994ADNDT}
 and those by DD-PC1~\cite{Niksic2008PR}, PC-F1 \cite{Burvenich2002PR}, PC-LA \cite{Nikolaus1992PR}, and NL3* \cite{Lalazissis2009PL}.
 The bold-faced quantities denote that the experimental values of the corresponding nuclei
 are used in the parametrization
 fitting. The root mean square (rms) deviation
 $\displaystyle\Delta=\sum\limits_i^N\sqrt{(r_i^{\rm exp}-r_i^{\rm cal})^2/N}$ and
 the root of relative square (rrs) deviation
 $\displaystyle\delta=\sum\limits_i^N\sqrt{\frac{(r_i^{\rm exp}-r_i^{\rm cal})^2/(r_i^{\rm exp})^2}{N}}$
 are respectively listed in the last two rows.}
 \label{Tabel:rch}
\begin{ruledtabular}
\begin{tabular}{ccccccc}
         Nuclei  &      Exp.  &          PC-PK1 &     DD-PC1     &   PC-F1        &      PC-LA     &       NL3* \\
\hline
    $^{16}\rm O$  &    2.737  &          2.7677 &    2.7472      &         2.7633 &\textbf{2.7528} &\textbf{2.7352} \\
    $^{40}\rm Ca$ &    3.4852 & \textbf{3.4815} &    3.4566      &\textbf{3.4777} &         3.4678 &\textbf{3.4704} \\
    $^{42}\rm Ca$ &    3.5125 & \textbf{3.4805} &    3.4626      &        3.4778  &         3.4729 &        3.4672  \\
    $^{44}\rm Ca$ &    3.5231 & \textbf{3.4826} &    3.4709      &        3.4809  &         3.4810 &        3.4672  \\
    $^{46}\rm Ca$ &    3.5022 & \textbf{3.4865} &    3.4806      &        3.4860  &         3.4912 &        3.4693  \\
    $^{48}\rm Ca$ &    3.4837 & \textbf{3.4890} &    3.4895      &\textbf{3.4906} &         3.5023 &\textbf{3.4705} \\
    $^{50}\rm Ti$ &     3.573 & \textbf{3.5558} &    3.5696      &        3.5664  &         3.5868 &        3.5442 \\
    $^{58}\rm Ni$ &    3.7827 &          3.7372 &    3.7761      &\textbf{3.7645} &         3.8065 &        3.7399 \\
    $^{88}\rm Sr$ &    4.2036 & \textbf{4.2247} &    4.2231      &\textbf{4.2269} &\textbf{4.2379} &        4.2159 \\
    $^{90}\rm Zr$ &    4.2720 & \textbf{4.2695} &    4.2664      &\textbf{4.2724} &         4.2847 &\textbf{4.2636} \\
    $^{92}\rm Mo$ &    4.3170 & \textbf{4.3125} &    4.3140      &         4.3192 &         4.3333 &        4.3087 \\
   $^{112}\rm Sn$ &    4.5957 &          4.5801 &    4.5894      &\textbf{4.5870} &         4.6044 &        4.5753 \\
   $^{116}\rm Sn$ &    4.6257 &          4.6121 &    4.6174      &         4.6168 &         4.6307 &\textbf{4.6039}\\
   $^{122}\rm Sn$ &    4.6633 & \textbf{4.6561} &    4.6579      &         4.6549 &         4.6728 &       {4.6430}\\
   $^{124}\rm Sn$ &    4.6739 & \textbf{4.6694} &    4.6714      &\textbf{4.6677} &         4.6864 &\textbf{4.6554} \\
   $^{138}\rm Ba$ &    4.8348 & \textbf{4.8508} &    4.8511      &         4.8494 &         4.8667 &        4.8369 \\
   $^{140}\rm Ce$ &    4.8774 & \textbf{4.8879} &    4.8879      &         4.8871 &         4.9037 &        4.8748 \\
   $^{144}\rm Sm$ &    4.9525 & \textbf{4.9544} &    4.9521      &         4.9547 &         4.9676 &        4.9484 \\
   $^{202}\rm Pb$ &    5.4772 &          5.4908 &    5.4869      &\textbf{5.4892} &         5.4996 &        5.4825 \\
   $^{204}\rm Pb$ &    5.4861 & \textbf{5.5005} &    5.4962      &         5.4987 &         5.5112 &\textbf{5.4916} \\
   $^{206}\rm Pb$ &    5.4946 & \textbf{5.5098} &    5.5049      &         5.5078 &         5.5200 &        5.5004 \\
   $^{208}\rm Pb$ &    5.5046 & \textbf{5.5185} &    5.5129      &\textbf{5.5162} &\textbf{5.5279} &\textbf{5.5087} \\
   $^{214}\rm Pb$ &    5.5622 &          5.5798 &    5.5711      &\textbf{5.5762} &         5.5813 &\textbf{5.5699} \\
\hline
   $\Delta$       &           &           0.019 &    0.019       &   0.017        &          0.023 &        0.022 \\
   $\delta$       &           &          0.53\% &    0.51\%      &  0.45\%        &         0.55\% &        0.60\% \\
\end{tabular}
\end{ruledtabular}
\end{table}

\begin{table}[!htbp]
\caption{The predicted saturation properties for nuclear matter by PC-PK1 in comparison with
those by DD-PC1, PC-F1, PC-LA, NL3*, and PK1.}
\label{Tabel:nuclearmatter}
\begin{ruledtabular}
\begin{tabular}{ccccccc}
                         &  PC-PK1 & DD-PC1  &   PC-F1 &   PC-LA &    NL3*  &        PK1 \\
\hline
$\rho_0$ $\rm (fm^{-3})$ &  0.154  &  0.152  &   0.151 &   0.148 &    0.150 &      0.148 \\

 $E/A \rm (MeV)$         &  -16.12 &  -16.06 &  -16.17 &  -16.13 &   -16.31 &     -16.27  \\

  $M_D^\ast/M$           &  0.59   &   0.58  &    0.61 &    0.58 &     0.59 &       0.60 \\

  $M_L^\ast/M$           &  0.65   &   0.64  &    0.67 &    0.64 &     0.65 &       0.66 \\

 $K_0 \rm(MeV)$          &  238    &   230   &     255 &     264 &      258 &        283 \\

$E_{\rm sym} \rm(MeV)$   &  35.6   &    33   &    37.8 &    37.2 &     38.7 &       37.6 \\

$L \rm(MeV)$             &  113    &    70   &     117 &     108 &      123 &        116 \\

$K_{\rm asy} \rm(MeV)$   &  -583   &    -528 &    -627 &    -709 &     -630 &       -641  \\
\end{tabular}
\end{ruledtabular}
\end{table}



%\end{CJK*}
\end{document}